\documentclass[12pt,preprint]{aastex}
\usepackage{appendix}
\usepackage{lscape}

\slugcomment{In memory of Paola D'Alessio, who is dearly missed.}

\begin{document}{}

\def\t0{\theta_{\circ}}
\def\muo{\mu_{\circ}}
\def\sd{\partial}
\def\be{\begin{equation}}
\def\en{\end{equation}}
\def\bv{\bf v}
\def\bvo{\bf v_{\circ}}
\def\ro{r_{\circ}}
\def\rhoo{\rho_{\circ}}
\def\etal{et al.\ }
\def\msun{\,M_{\sun}}
\def\rsun{\,R_{\sun}}
\def\rstar{\,R_{\star}}
\def\lsun{L_{\sun}}
\def\msunyr{M_{\sun} yr^{-1}}
\def\kms{\rm \, km \, s^{-1}}
\def\mdot{\dot{M}}
\def\mdotd{\dot{M}_{\rm d}}
\def\Md{\dot{M}}
\def\curf{{\cal F}}
\def\ecs{erg cm^{-2} s^{-1}}
\def \haebe{HAeBe}
\def \mum {\,{\rm \mu m}}
\def \simali {{\sim\,}}
\def \K {\,{\rm K}}
\def \Angstrom     {\,{\rm \AA}}
\newcommand \g            {\,{\rm g}}
\newcommand \cm           {\,{\rm cm}}

\title{Emission from Water Vapor and Absorption from Other Gases at 5--7.5 Microns 
in {\it Spitzer}-IRS Spectra of Protoplanetary Disks}

\author{B.~A. Sargent\altaffilmark{1,2},
W. Forrest\altaffilmark{3},
Dan M. Watson\altaffilmark{3},
N. Calvet\altaffilmark{4},
E. Furlan\altaffilmark{5,6},
K.~H. Kim\altaffilmark{3,7},
J. Green\altaffilmark{8},
K. Pontoppidan\altaffilmark{2},
I. Richter\altaffilmark{3},
C. Tayrien\altaffilmark{3,9}
}
\altaffiltext{1}{Center for Imaging Science and Laboratory for Multiwavelength 
                 Astrophysics, Rochester Institute of Technology, 54 Lomb Memorial Drive, 
                 Rochester, NY 14623, USA; {\sf baspci@rit.edu}}
\altaffiltext{2}{Space Telescope Science Institute, 3700 San Martin 
                 Drive, Baltimore, MD 21218, USA}
\altaffiltext{3}{Department of Physics and Astronomy, University of 
                 Rochester, Rochester, NY 14627}
\altaffiltext{4}{Department of Astronomy, The University of Michigan, 500 Church 
                 Street, 830 Dennison Building, Ann Arbor, MI 48109, USA}
\altaffiltext{5}{National Optical Astronomy Observatory, 950 North Cherry Avenue, 
                 Tucson, AZ 85719, USA}
\altaffiltext{6}{Visitor at the Infrared Processing and Analysis Center, 
                 Caltech, 770 South Wilson Avenue, Pasadena, CA 91125, USA}
\altaffiltext{7}{Korea Astronomy and Space Science Institute, 776, Daedeokdae-ro, 
                 Yuseong-gu, Daejeon, 305-348, Republic of Korea}
\altaffiltext{8}{Department of Astronomy, University of Texas, 1 University Station, 
                 Austin, TX 78712, USA}
\altaffiltext{9}{Department of Computer Science, Rochester Institute of Technology, 
                 102 Lomb Memorial Drive, Rochester, NY 14623, USA}
\begin{abstract}

We present spectra of 13 T Tauri stars in the Taurus-Auriga 
star-forming region showing emission in {\it Spitzer 
Space Telescope} Infrared Spectrograph (IRS) 5--7.5$\mu$m 
spectra from water vapor and absorption from other gases in these 
stars' protoplanetary disks.  Seven stars' spectra show an 
emission feature at 6.6$\mu$m due to the $\nu_{2}$ = 1--0 bending 
mode of water vapor, with the shape of the spectrum suggesting 
water vapor temperatures $>$500\,K, though some of these spectra also 
show indications of an absorption band, likely from another molecule.  This 
water vapor emission contrasts with the absorption from warm water vapor seen in 
the spectrum of the FU Orionis star V1057 Cyg.  The 
other six of the thirteen stars have spectra showing a strong absorption band, 
peaking in strength at 5.6--5.7$\mu$m, which for some is consistent with gaseous 
formaldehyde (H$_{2}$CO) and for others is consistent with gaseous formic acid (HCOOH).  
There are indications that some of these six stars may also have weak water 
vapor emission.  Modeling of these stars' spectra suggests these 
gases are present in the inner few AU of their host disks, consistent 
with recent studies of infrared spectra showing gas in protoplanetary 
disks.

\end{abstract}


\section{Introduction}

Water has been observed 
in its gaseous form in numerous astrophysical environments.  Its 
signature is seen in spectra of stellar atmospheres 
\citep{jor01,jones02,roe04,cush06}, the extended atmosphere of 
Mira \citep{kuiper62,yama99}, and the circumstellar realms of numerous 
oxygen-rich Asymptotic Giant Branch stars \citep{cami02,sarg10}.  
The absorption and emission lines attributed to 
water in the 5.5--7.2$\mu$m Becklin-Neugebauer/Kleinman-Low (BN/KL) spectrum are identified as 
belonging to the $\nu_{2}$ = 1--0 bending mode of water vapor \citep[][]{gonzal98}.  
Spectral lines from water vapor were observed in spectra of the 
molecular cloud cores surrounding 
embedded Young Stellar Objects \citep[YSOs;][]{helm96,vd96,dar98}.  
Water vapor emission lines have been seen in the {\it Spitzer Space Telescope} 
\citep{wer04} Infrared Spectrograph \citep[IRS;][]{hou04} spectra of the 
protostar NGC 1333-IRAS 4B \citep{wat07}.

\citet{cg97} noted that one of the most important coolants in young 
circumstellar disks is water vapor, and they discuss how gas may 
regulate the temperatures of dust grains in these disks.  Observations of 
such protoplanetary disks are therefore crucial in helping to constrain the 
physics of such disks.  Emission from water vapor has been seen in 
near-infrared (1 $< \lambda/\mum <$ 2.3) spectra of circumstellar disks 
\citep{carr04} and in the far-infrared by Herschel-HIFI \citep[][]{hoger11}.  
\citet{cana08} presented a mid-infrared {\it 
Spitzer}-IRS spectrum of the classical T Tauri star (TTS) AA Tau from 
10--34 $\mu$m showing numerous emission lines from water vapor and the 
simple organic molecules HCN and C$_{2}$H$_{2}$.  Numerous other studies 
\citep[][]{mand08,sal08,pont10et,pont10,cana11,sal11,mand12,naj13} have shown 
spectra indicating emission from these and other gases suggesting an origin 
in the inner few AU of protoplanetary disks.  \citet{naj10} report HCO+ and 
possibly CH$_{3}$ from a {\it Spitzer}-IRS spectrum of TW Hya.  These studies 
have typically made use of high-resolution {\it Spitzer}-IRS spectra; however, 
low-resolution {\it Spitzer}-IRS spectra have also been used to study molecular 
(including simple organic) emission \citep[see studies of HCN and C$_{2}$H$_{2}$ 
by][]{pas09,teske11}.  Here we report the detection of emission from water vapor 
(H$_{2}$O) and absorption from the simple organic molecules formaldehyde 
(H$_{2}$CO) and formic acid (HCOOH) in the inner regions 
of a number of TTSs in the Taurus-Auriga star-forming region as revealed 
by their 5--7.5$\mu$m spectra.

\section{Data}

\subsection{Spectral Data Reduction}

We analyze the spectra of 13 TTSs in the Taurus-Auriga star-forming region 
(Table 1).  We selected the sources with the strongest emission or absorption 
structures in the 5--7.5 $\mu$m range not due to previously identified 
gas or solid-state species (e.g. polycyclic aromatic hydrocarbons -- PAHs) 
or ices \citep[][]{pont05,fur08,zas09,boog11} from the \citet{fur06b} sample.  
All 5--7.5$\mu$m spectra included in our analysis were reduced and prepared 
by the same techniques (including dereddening) as described by \citet{fur06b} 
and \citet{sarg09b}, including deriving uncertainties from half the difference between 
the spectra obtained at the 2 nod positions, and applying a 1\% lower threshold to 
the relative uncertainties, such that any relative uncertainty less than 1\% was set 
to 1\%.  The spectra treated this way includes the spectrum of V1057 Cyg, a FU Orionis 
object whose spectrum was shown by \citet{green06} and suggested by those 
authors to have water vapor absorption bands over 5--7.5 $\mu$m wavelengths.  
Other spectra reduced and prepared this way include the Class III YSOs, HBC 388 
and LkCa 1 \citep[][]{fur06b}, which were chosen because they have no dust emission 
in excess of 2\% of the continuum over 5--35 $\mu$m (i.e., the level of the flat field 
variation of the IRS; see the {\it Spitzer}-IRS 
Handbook\footnote{http://irsa.ipac.caltech.edu/data/SPITZER/docs/irs/irsinstrumenthandbook/home/}), 
so any structure seen in their spectra must be photospheric in origin.  As such, 
they are useful to compare to our TTS sample of 13 spectra.

\subsection{Spectra with the 6.6$\mu$m Emission Feature}

The spectra of CI Tau, DF Tau, DR Tau, FS Tau, FZ Tau, GN Tau, and 
RW Aur A each show what appears to be an emission feature that peaks at 
$\simali$6.6$\mu$m.  Figure 1 shows this feature for RW Aur A to be much larger in 
amplitude than the error bars on the data points composing the feature.  The 
same is found for the other 6 spectra showing strong 6.6 $\mu$m emission 
features in Figures 2 and 3.  In Table 1, we quantify the significance of the 6.6 
$\mu$m feature by providing the ratio of the absolute value of the equivalent 
width of the feature to its uncertainty.  For RW Aur A and the other 6 stars 
showing strong 6.6 $\mu$m emission features, this ratio is always greater than 8.  
In order to determine whether a feature was an artifact, we examined spectra 
of comparably bright TTSs, including DP Tau, AA Tau, FN Tau, GH Tau, and HK Tau 
\citep[see Fig. 3 of][]{fur06b}, to determine the consistency of the 5--7.5 $\mu$m 
spectral band.  The 5 comparison TTS spectra from \citet{fur06b} show hardly any 
spectral structure at all over these wavelengths; instead, one sees smooth continuum 
for these 5 spectra.  All of the 13 TTS spectra we analyze in our study plus the 5 
comparison TTS spectra from \citet{fur06b} have been reduced 
in a similar manner; e.g., variable column width aperture extraction, the 
use of RSRFs, etc.  Therefore, we believe the 6.6$\mu$m emission features 
to be intrinsic to the TTSs and not artifacts.

The lack of this feature in Class 
III YSO spectra (e.g., see the spectra of HBC 388 and Lk Ca 1 in Figure 5) 
further argues against these features being artifacts.  In general, the Short-Low 
detector of {\it Spitzer}-IRS (5--14 $\mu$m wavelength) is very well behaved, 
not suffering from the degree of the ``rogue'' pixel phenomenon \citep[e.g.,][]{wat07} 
that, for example, the Long-High detector suffers.  Finally, the degree of 
mispointing measured in the cross-dispersion direction (i.e., along the length 
of the slit) varies in our sample from less than 0.1 pixel width (the 
Short-Low pixel width is 1.8$\arcsec$) to 0.8 pixel width.  As shown by 
\citet{sargphd} for the {\it Spitzer}-IRS Mapping Mode observation of DH Tau, 
IQ Tau, and IT Tau (AOR \# 3536128), the dispersion-direction (perpendicular 
to the length of the slit) mispointing tracked with the cross-dispersion mispointing.  
Therefore, mispointing of the observations does not seem to be responsible 
for the features seen at these wavelengths.  It is also worth noting that some of these 
seven spectra showing strong 6.6 $\mu$m emission also have significant absorption 
in the 5.7 $\mu$m band.  This includes FZ Tau, GN Tau, RW Aur A, and especially 
DF Tau.  These four have 5.7 $\mu$m absorption band equivalent width to uncertainty 
ratios greater than 5 (see Table 1).

\subsection{Spectra with the 5.7 $\mu$m Absorption Band}

The spectra of DL Tau, FX Tau, HN Tau, IP Tau, IQ Tau, and V836 Tau each 
show what appears to be an absorption band centered around 5.7$\mu$m 
(Figures 4 and 5).  As with the emission features in Figures 1, 2, and 3, Figures 4 
and 5 show these absorption bands to be much larger in amplitude than the error bars 
on the data points composing the band.  We computed the significance of this 
feature in a manner similar to how it was done for the 6.6 $\mu$m emission feature 
and provide it also in Table 1.  Note that the significance of the 5.7 $\mu$m absorption 
band is always greater than 4.  Also, two of these 5.7 $\mu$m absorption band 
exemplars, HN Tau and IP Tau, have 6.6 $\mu$m emission feature equivalent width 
to uncertainty ratios greater than 3 (Table 1).  Thus, the presence of the 6.6 $\mu$m 
feature does not exclude the presence of the 5.7 $\mu$m band, and vice-versa.

\section{Analysis}

\subsection{Models of the Seven Spectra with Strong 6.6 $\mu$m Emission Features}

We argue that the 6.6 $\mu$m emission feature seen in the 5--7.5$\mu$m 
spectra of some of our sample arises from emission from water vapor, 
H$_{2}$O.  Specifically, the water vapor emission seen over these wavelengths 
comes largely from the $\nu_{2}$ = 1--0 bending mode of water vapor, as in the case of 
the features seen by \citet{gonzal98} in the spectrum of the BN/KL region of Orion.  
In the \citet{gonzal98} BN/KL spectrum, the R branch ($\lambda < 6.3 \mu$m) lines 
are seen in absorption, the P branch ($\lambda > 6.3 \mu$m) lines are seen in 
emission; of the Q branch lines, some are in absorption, while others are in emission.

In Figure 1, we plot the spectrum and model of RW Aur A.  We begin with RW Aur A 
to identify the most likely origin of the 6.6 $\mu$m feature.  Our model includes water 
vapor only in emission.  Full radiative transfer modeling is beyond the scope of the 
present work.  We assume LTE for our models, and for RW Aur A, we assume a model with three 
components added together.  We vary the temperature, column density, and 
inferred radius of emitting region to match the model to the data, judging the quality of 
fit by eye.  We kept the model as simple as possible, but sufficiently complex so that 
the model fits the data well.  The intensity of radiation emerging from each 
component is of the form

\begin{eqnarray}
I_{\nu} & = & I_{\nu, 0} e^{-\tau_\nu} + S_{\nu}(1-e^{-\tau_\nu})
\end{eqnarray}

\noindent where $I_{\nu, 0}$ is the background intensity, $\tau_{\nu}$ is the optical depth 
of the slab of gas, and $S_{\nu}$ is the source function of the slab of gas (i.e., a Planck 
function at the temperature of the gas).  We assumed the water vapor to have a microturbulent 
velocity of 1 km s$^{-1}$.  Because of the low resolution of these {\it Spitzer}-IRS spectra, we have 
no information on the widths of the individual lines, as the water lines are not resolved, but 
our assumed microturbulent velocity is similar to the $\sigma$ = 2 km s$^{-1}$ local line width assumed 
by \citet{sal08} for their gas modeling of {\it Spitzer}-IRS spectra of protoplanetary disks.  
The temperature of the water vapor giving rise to the 6.6 $\mu$m feature cannot be 
much lower than 500\,K, or else the 6.6 $\mu$m feature would become too weak to match the 
spectrum.  We used the HITEMP2010 line list for the main isotopologue of water vapor 
\citep{roth10}, and the HITRAN2012 line list for the main isotopologue of formaldehyde 
\citep{roth13}.  For the partition function of these molecules at lower temperatures 
(T $\leq$ 75\,K), we use the partition functions available at the Jet Propulsion Laboratory 
website\footnote{http://spec.jpl.nasa.gov/ftp/pub/catalog/catdir.cat} \citep[][]{pick98}.  For 
partition functions at higher temperatures, we use HITRAN2008 \citep[][]{roth09}.  We 
then rebin the model to the spectral resolution of the IRS (R $\sim$ 60-120) over these 
wavelengths.

For RW Aur A, our model is a sum of the flux from 3 components, each of the form 
F$_{\nu,i}$ = $\Omega_{i}$I$_{\nu,i}$, where the intensity, I$_{\nu,i}$, for each is 
determined using Equation 1 (though some of the models use this equation iteratively 
to determine the emergent intensity for a more complicated geometry; see Section 3.2), 
each component with its own independent solid angle.  
The first component is an isothermal slab of water vapor at 1100\,K of column 
density 1.7$\times$10$^{15}$ cm$^{-2}$, with zero background intensity.  The second 
component is a naked blackbody (i.e., no slab in front of it, so $\tau_{\nu}$ = 0) at 190\,K.  
Both these two components have the same solid angle, which is equal to that of a face-on 
disk of radius 920$\rsun$ at a distance of 139pc \citep[the assumed distance to the 
Taurus/Auriga star-forming region; here, we follow][]{bert99}.  The third component 
features a 1400\,K blackbody behind an isothermal cloud of formaldehyde (H$_{2}$CO) 
at 500\,K of column density 7$\times$10$^{18}$ cm$^{-2}$.  The solid angle for the third 
component is equivalent to that of a face-on disk of radius 16.1$\rsun$ at 139 pc.  Though 
the solid angle for the water emission is much greater than that for the H$_{2}$CO 
absorption, the actual abundance ratio of the two (Table 2) is roughly 8:1 in favor of 
water.  Perhaps the water giving rise to the spectral emission we see originates from a 
wider range of disk radii, while the formaldehyde originates from a narrower range.  
Because we do not perform self-consistent radiative transfer modeling of the gas 
emission and absorption, we cannot say much more about the relative locations in the 
protoplanetary disks of the water and the formaldehyde.  The 
assumed column densities of water and formaldehyde are similar to or lower than the 
values for the molecules included in the model of AA Tau by \citet{cana08} for wavelengths 
longward of 10 $\mu$m.  Also, the solid angles of these components 
suggest regions in the inner few AU of the protoplanetary disk, consistent with the 
water vapor inferred from modeling by \citet{cana08}.  That our model produces a 6.6$\mu$m 
emission feature means there must be a temperature inversion: the water vapor is hotter 
than the material underneath, further supporting an origin in the cooler disk regions in 
the inner few AU.  The maximum optical depth of any of the H$_{2}$O lines in 
the model of RW Aur A over 5--7.5$\mu$m is $\sim$ 0.01, while the maximum optical 
depth of any of the H$_{2}$CO lines in the model is $\sim$ 4.4.

The model matches the 6.6 $\mu$m feature quite well, including the 
``shoulder'' on the feature at 6.5 $\mu$m.  Further, our model matches the overall shape 
of the 5--7.5 $\mu$m spectrum quite well, with only a few excursions outside of the 
spectral error bars.  The water vapor emission in our model also matches the minor 
emission bump seen at 6.85 $\mu$m in the spectrum.  We caution, however, that the 
``features'' seen at 6.6 $\mu$m, 6.85 $\mu$m, etc are actually manifolds of multitudes of 
narrow water vapor lines that {\it Spitzer}-IRS cannot resolve at this low spectral resolution.  
For example, between 6.3 and 6.75 $\mu$m in the RW Aur A model (roughly corresponding 
to the 6.6 $\mu$m ``feature'', or manifold), there are $>$ 500 lines of peak flux $>$ 0.05 Jy 
in the unconvolved model.  Much of the water vapor emission seen over 5--7.5 $\mu$m 
belongs to the water vapor $\nu_{2}$ = 1--0 bending mode that \citet{gonzal98} identified 
in their spectrum of BN/KL.  Between 5.6--6.1 $\mu$m, the model is mostly within the error 
bars of the spectrum.  The excursions over this range may be due to the simplicity of assuming 
absorption from formaldehyde at only one temperature (500\,K).  For our model of RW Aur A, we 
assume an abundance ratio of water to formaldehyde of $\sim$ 8 (Table 2).  A range of 
temperatures of formaldehyde (see the models of DL Tau, HN Tau, and IQ Tau in Section 
3.2) may improve the match of the model to the spectrum over this wavelength range.  

As further verification of the identity of the species giving rise to the 6.6 $\mu$m feature, 
we model the spectrum of the FU Orionis object, V1057 Cyg.  This star's {\it Spitzer}-IRS 
spectrum was first presented by \citet{green06}, who noted the presence of weak 
absorption bands between 5--7.5 $\mu$m in many of the FU Orionis type objects' IRS 
spectra, including that of V1057 Cyg, and attributed these bands to water vapor.  Indeed, 
\citet{green13} found emission from water in a Herschel-PACS spectrum of V1057 
Cyg.  Following the suggestion of \citet{green06}, we constructed a model of water vapor 
absorption for this object's spectrum.  The first model component is an isothermal cloud of 
water vapor at 800\,K of column density 6$\times$10$^{19}$ cm$^{-2}$ in front of a 950\,K 
blackbody subtending a solid angle equivalent to a face-on disk of radius 227 $\rsun$ at 
a distance of 600pc \citep[][]{hk96}, while the second model component is merely a 200\,K 
blackbody subtending a solid angle equivalent to that of a face-on disk of radius 1800 
$\rsun$ at a distance of 600pc \citep[][]{hk96}.  For model details, see Table 2.  The 
microturbulent velocity assumed for the 
water lines in the V1057 Cyg model is 10 km/s.  The quality of the fit of the model to the 
V1057 Cyg spectrum is quite remarkable, considering the simplicity of the model.  The 
model also matches the manifolds between 5.2--5.8 $\mu$m. In a number of 
ways the V1057 Cyg spectrum is a mirror of the RW Aur A spectrum: V1057 Cyg (RW 
Aur A) has a local maximum (minimum) at 6.3 $\mu$m, a shoulder at 6.5 $\mu$m, a 
local minimum (maximum) at 6.6 $\mu$m, a local minimum (maximum) at 6.8 $\mu$m, 
etc.   One notable difference, however, is that the overall shape of the H$_{2}$O 
absorption band in the V1057 Cyg spectrum between 5.2--6.3 $\mu$m is not mirrored 
in a pure emission band in the RW Aur A spectrum.  For comparison, a {\it Spitzer}-IRS 
spectrum of a pure H$_{2}$O emission band over these wavelengths can be seen in 
the spectrum of the lower-mass Oxygen-rich AGB star shown by \citet{sarg10}.  This 
supports our claim that there is absorption present in the RW Aur A spectrum at these 
wavelengths, and that formaldehyde is a good candidate to explain this.

We modeled the 6 other spectra showing strong 6.6 $\mu$m emission features in a 
manner similar to how we modeled RW Aur.  The spectra and their best-fit models are 
shown in Figures 2 and 3, and the model parameters are listed in Table 2.  We conclude 
water vapor emission is responsible for the 6.6 $\mu$m features seen in the observed 
spectra of RW Aur A (Figure 1) and the 6 other TTSs plotted in Figures 2 and 3.

For the spectra of FZ Tau, FS Tau, and GN Tau, we used models similar to that of RW 
Aur A, in that we used water and formaldehyde (Figure 2).  As can be seen, the models 
match the observed data fairly successfully, matching the 6.6 $\mu$m feature well.  The 
model of FZ Tau additionally successfully matches the $\sim$ 0.1 $\mu$m wide manifolds 
at 6.85, 5.7, 5.9, and 6.05 $\mu$m.  Overall, the manifolds are weaker for FS Tau, and its 
model reflects that.  Considering that the higher relative uncertainties on the data for GN 
Tau, the model fit is still satisfactory.

The spectra of DR Tau, DF Tau, and CI Tau require absorption from a molecule other 
than formaldehyde, as formaldehyde produces an absorption band too wide for these 
spectra, especially for DF Tau (Figure 3).  Instead, for the models of these spectra, we 
replace formaldehyde with formic acid \citep[HCOOH; line list for the main isotopologue 
from HITRAN2008; see][]{roth09}.  Again, the models match the spectra fairly successfully.  
For DF Tau, the formic acid absorption feature matches the absorption band very well in 
strength, width, central wavelength.  It even matches the ``dimple'' at the bottom of the 
feature, around 5.65 $\mu$m.  The model of DR Tau mostly matches the manifolds seen 
in its spectrum.  The model of CI Tau also matches its spectrum well, though the relative 
uncertainties are somewhat higher than the other spectra in Figure 3.

\subsection{Models of the Six Spectra with Strong 5.7 $\mu$m Absorption Bands}

The other 6 TTS spectra in our sample of 13 show either weak 6.6 $\mu$m emission 
features or none at all.  Rather, they show strong absorption bands centered at 5.7 
$\mu$m.  We note that these absorption bands are centered at too short a wavelength 
to be consistent with water ice absorption \citep[see][]{pont05}.  Organic ice mixtures 
are another possibility to consider.  Such ices involve mixtures of H$_{2}$O, CH$_{3}$OH, 
NH$_{3}$, CO, CH$_{4}$, H$_{2}$CO, and other molecules, in varying ratios, both without 
and with ultraviolet photolysis, at varying temperatures 
\citep[][]{greenberg86,allaman88,schutte93,bern94,bern95,gera96,munoz03,nuevo06,bis07,danger13}.  
The spectra of these organic ice mixtures are similar in that, though 
features begin appearing around 5.7 $\mu$m, the features grow stronger toward 10 
$\mu$m wavelength.  The same is true for kerogen and Murchison meteorite organic 
residue \citep[][]{khare90}.  The spectrum of H$_{2}$CO:NH$_{3}$ ice at 10\,K after 
deposition shown by \citet{schutte93} is somewhat different 
in that its 5.8 $\mu$m feature is stronger than the other features out to 10 $\mu$m; 
however, the 5.8 $\mu$m does not match the central wavelengths of the absorption 
bands we observe in our TTS sample.  Such behavior observed in the infrared spectra 
of all these studies of organic solids and ice mixtures is not observed in our spectra, 
so we do not attribute the 5.7 $\mu$m absorption in our TTS spectra to such organic solids 
and ice mixtures.

These 5.7 $\mu$m absorption bands are clearly seen in the spectra 
of DL Tau, HN Tau, IQ Tau, FX Tau, IP Tau, and V836 Tau (Figures 4 and 5) and 
are also seen for some of the stars showing the 6.6 $\mu$m emission feature 
(see previous subsection).  For the model of RW Aur A (Figure 1 and Section 3.1), 
we modeled this absorption by including formaldehyde, H$_{2}$CO, in our model, 
in addition to water vapor.  We do the same for HN Tau (Figure 4), as it seems to 
have a weak 6.6 $\mu$m emission feature.  However, the 5--7.5 $\mu$m spectra 
of the other 5 spectra in Figures 4 and 5 seem to require no water vapor in their 
models.

Figure 4 shows the spectra and models of DL Tau, HN Tau, and IQ Tau.  These models 
of DL Tau, HN Tau, and IQ Tau assume there is a layer of cold formaldehyde in front 
of a layer of warmer formaldehyde, in front of a hotter blackbody, and that the solid 
angle of the layers and blackbody are the same (Table 2).  The model for HN Tau 
additionally includes a separate component of water vapor emission.  In order to compute 
the part of the emergent spectrum giving rise to the absorption band, Equation 1 is used 
iteratively, such that the first iteration determines 
intensity I$_{\nu}$ from radiation emerging from the first layer, where the hot blackbody 
is I$_{\nu, 0}$, the source function of the lower (warmer) formaldehyde layer is S$_{\nu}$, 
and $\tau_{\nu}$ is the optical depth of the lower formaldehyde layer.  The intensity 
determined from this first iteration, I$_{\nu}$, becomes the background intensity, 
I$_{\nu, 0}$, for the second iteration of Equation 1, where S$_{\nu}$ in the second 
iteration is the source function for the gas in the upper (cooler) formaldehyde layer, and 
$\tau_{\nu}$ in the second iteration is the optical depth of this upper formaldehyde layer.  
The warmer formaldehyde provides much of the absorption over the entire band, though 
less so at the center.  The cooler formaldehyde provides the rest of the absorption at the 
center of the band.  The maximum optical depth of any of the lines of formaldehyde over 
5--7.5 $\mu$m comprising this model is $\simali$11.  
As noted previously, the absorption bands seen in our sample at these wavelengths 
seem to break down into two types: a wider band and a narrower band.  The three 
spectra shown in Figure 4 have the wider band and are well modeled by 
formaldehyde.  This wider band typically spans the region 5.4--6.0 $\mu$m.

The spectra of FX Tau, IP Tau, and V836 Tau, shown in Figure 5, have the narrower 
absorption band that is too narrow to be fit by formaldehyde, which fit the wider bands 
shown in Figure 4.  This narrow band typically spans about 5.45--5.85 $\mu$m.  For 
these 3 spectra, we assume absorption from formic acid instead of formaldehyde.  As 
with the models in Figure 3, the formic acid produces a narrower band in each model 
shown in Figure 5 that mostly matches the observed absorption features.  The best 
match is for V836 Tau, in terms of the width, strength, and central wavelength of the 
band, though the relative uncertainties are somewhat large.  For IP Tau, the strength 
of the band is well modeled, though the model band is a little narrower and shifted 
slightly to longer wavelengths than the band seen in the data.  The same is true for 
FX Tau, with a discrepancy between model and data perhaps slightly greater than 
was true for IP Tau.

We also include in Figure 5 the spectra of two Class III YSOs without any dust 
excess.  These spectra - HBC 388, a K1 star, and LkCa 1, an M4 star - demonstrate 
that the photospheric contribution to any features seen at 5--7.5 $\mu$m is very 
minor, if present at all.  The spectral types of these stars span most of the range of 
the spectral types for our sample of TTS (Table 1).  HBC 388 shows hardly any 
absorption bands over these wavelengths at all.  LkCa 1 shows a very shallow 
absorption band at 6.3--7 $\mu$m that is possibly due to water vapor in the stellar 
photosphere \citep[][]{roe04,cush06}.  However, excess infrared emission for our sample 
of TTSs essentially ``fills in'' such bands in the stellar photosphere spectra of our TTSs.  
The ratios of the observed flux at 6.3 $\mu$m (which, relatively free from water vapor 
lines, should give the best indication of the local dust continuum) to the inferred stellar 
photosphere flux \citep[interpolated at this wavelength from the photospheric SEDs 
plotted for our TTS sample by][]{fur06b} at this wavelength, $\beta_{6.3}$, are all very 
high, between 3.8 (FX Tau) and 31.8 (HN Tau; see Figures 4 and 5).  Thus, the emission 
and absorption features and bands that we model in our TTS spectra are circumstellar 
in origin, not photospheric.

\subsection{Equivalent Widths, Central Wavelengths, and Integrated Fluxes of the 5.7 $\mu$m Band 
and 6.6 $\mu$m Feature for Taurus-Auriga TTSs}

We selected our sample of 13 TTSs from Taurus/Auriga because they showed 
the strongest 6.6 or 5.7 $\mu$m features.  However, these features are seen in 
other TTS spectra from the Taurus/Auriga star-forming region.  It is beyond the 
scope of the present study to model the emission seen in all such stars.  However, 
it is relatively simple to measure the equivalent widths, central wavelengths 
\citep[defined in the same sense as they were defined by][]{sloan07}, and integrated fluxes
for our spectra.  Equivalent widths provide a relative measure the strength of the emission or 
absorption features, while integrated fluxes provide an absolute measure of such strengths.  
The central wavelengths are important, especially for the 5.7 $\mu$m 
band because the central wavelength of the feature shifts according to the type 
of molecule providing the absorption (shorter central wavelength for formic acid, 
longer central wavelength for formaldehyde).  Our modeling work informs these 
measurements by suggesting the wavelength ranges to fit continua and to measure 
central wavelength, equivalent width, or integrated flux.

We began by determining continua for the 6.6 and 5.7 $\mu$m features.  We fit a 
power law continuum of the form F$_{\nu,mod}$ = 10$^{C}\lambda^{P}$ to the data 
just outside each feature by least-squares; we used 5.36--5.429 and 6.026--6.10 $\mu$m 
for the 5.7 $\mu$m band and 6.265--6.34 and 6.7--7.0 $\mu$m for the 6.6 $\mu$m band.  
After determining the continuum to use for each feature, the equivalent width and integrated 
flux (defined as the difference between observed spectrum and computed continuum, 
integrated over the band) was computed over 5.428--6.026 $\mu$m for the 5.7 $\mu$m 
band and 6.34--6.7 $\mu$m for the 6.6 $\mu$m feature.  We computed these equivalent 
widths and integrated fluxes for all the TTSs 
analyzed by \citet{sarg09b} except for 4 -- CoKu Tau/4, DM Tau, GM Aur, and Lk Ca 15 -- 
as these TTSs have inner holes in their disks, so that there is little to no disk emission 
over 5--7.5 $\mu$m wavelength.

Next, we measured the central wavelengths of these features.  To do this, we integrated, 
channel-by-channel across the features over the same wavelength ranges used to 
measure equivalent widths, from each end towards the other until we reached the 
first channel that put the channel integration sum over 50\% of the total integrated 
flux within the feature.  We then interpolated between this channel and the previous one 
to determine the wavelength corresponding to exactly 50\% integrated flux.  This was 
performed in both increasing and decreasing wavelength directions.  The central wavelength 
is then the average of the result obtained from the two different directions.

We then searched for trends between the 5.7 and 6.6 $\mu$m central wavelengths and 
equivalent widths and the dust model parameters, stellar properties, and disk properties 
explored by \citet{sarg09b}, and also with the 5.7 and 6.6 $\mu$m central wavelengths, 
equivalent widths, and integrated fluxes themselves.  The three H$_{2}$CO absorption exemplars (DL Tau, 
HN Tau, and IQ Tau; see Figure 4) do have 5.7 $\mu$m absorption band central 
wavelengths (5.668, 5.733, and 5.728 $\mu$m, respectively; see Table 1) systematically 
long-ward of the central wavelengths for the same band in the 3 HCOOH absorption 
exemplars (FX Tau, 5.560 $\mu$m; IP Tau, 5.597 $\mu$m; and V836 Tau, 5.656 $\mu$m; 
Figure 5; Table 1).  However, very little in the way of significant correlations were found.  
Figure 6 shows the 5.7 $\mu$m equivalent width versus the 5.7 $\mu$m central wavelength.  
For smaller equivalent widths, there is a wide range of central wavelengths, but for the 
larger equivalent widths, the central wavelength ranges only between about 5.6--5.7 
$\mu$m, which is consistent with the modeling described in Sections 3.1 and 3.2.  
We computed the correlation coefficient, r, and the probability of 
a correlation coefficient of equal or greater magnitude being found for a non correlated 
data set \citep[][]{taylor82}, P, in the same manner as \citet{sarg09b}.  We found r = 0.18 and 
P = 16\% for the pair of 5.7 $\mu$m equivalent width and 5.7 $\mu$m central wavelength, 
so there is no correlation between this pair.  We also searched for a trend between 
the 5.7 $\mu$m and 6.6 $\mu$m features' equivalent widths.  These are plotted in 
Figure 7.  For this pair, we also find no correlation, with r = 0.24 and P = 6\%.

One pair that illustrates the typical sort of weak trend found between an equivalent width or 
central wavelength is between the spectral continuum index, n$_{13-31}$, measuring the 
continuum color in the infrared spectrum \citep[][]{fur11}, and the 5.7 $\mu$m central 
wavelength.  This is plotted in Figure 8, with r = -0.31 and P = 2\%, indicating there is a very weak indication of 
more negative 13-to-31 micron spectral index (bluer color) with increasing 5.7 $\mu$m 
central wavelength.  However, at a glance, little correlation of note appears in Figure 8.  More 
negative n$_{13-31}$ index suggests a more settled disk \citep[][]{fur11}.  The trend would 
thus suggest that more settled disks tend to have the absorption band shifted to longer 
wavelengths, presumably because of greater H$_{2}$CO abundance relative to HCOOH.  
This potential trend should be explored in a larger sample of TTS spectra.  
Caution, however, is urged in attempting to interpret a correlation between 
n$_{13-31}$ index and 5.7 $\mu$m central wavelength.

\subsection{Comparison of 6.6 $\mu$m Ro-vibrational and 17 $\mu$m Ground State 
H$_{2}$O Emission}

Of the seven stars with strong H$_{2}$O emission (see Section 3.1), \citet{naj13} report $\sim$ 17 
$\mu$m H$_{2}$O line fluxes from {\it Spitzer}-IRS Short-High ($\sim$ 10--19 $\mu$m) 
spectra for four: CI Tau, DR Tau, FZ Tau, and RW Aur A.  By widening the comparison to all 
stars in common between the \citet{naj13} sample and all stars in our sample for which we 
measure equivalent widths and integrated fluxes, we obtain a sample of 23 stars.  We 
plot in Figure 9 the 17 $\mu$m H$_{2}$O line fluxes from \citet{naj13} versus the 6.6 $\mu$m 
integrated fluxes we measure.  There is a definite correlation between the \citet{naj13} 17 $\mu$m 
H$_{2}$O line flux with our 6.6 $\mu$m integrated fluxes of r = 0.79 and P = 0.1 \%.

\citet{cana11} note that H$_{2}$O lines in the 16--18 $\mu$m region are high-excitation 
ground state lines.  \citet{finzi77} found that water 
relaxes from its stretching modes, $\nu_{1}$ and $\nu_{3}$, by first transferring to the 
bending overtone, $\nu_{2}$ = 2, then to the bending fundamental, $\nu_{2}$ = 1, then 
to the ground state; thus, the bending mode of water is strongly coupled to the rotational 
levels in the ground state \citep[see also][]{gonzalf99}.  The correlation between the rovibrational 
5--7.5 $\mu$m H$_{2}$O emission and the 17 $\mu$m H$_{2}$O ground state lines could be 
consistent with non-LTE excitation of these H$_{2}$O ground state lines by radiative pumping through 
the H$_{2}$O $\nu_{2}$ = 1 mode.  As \citet{mand12} 
summarize, modeling by \citet{meij09} and observations by \citet{mand08} and \citet{pont10et} 
suggest radiative pumping from the star enhances rovibrational H$_{2}$O emission from 
protoplanetary disks.  We urge caution, however, as this sample size is somewhat small (23).  
In the future, we will compare the 6.6 $\mu$m H$_{2}$O feature to the 17 $\mu$m H$_{2}$O 
line fluxes in a larger sample.

\section{Discussion}

We note the temperatures required ($>$500\,K) to fit the 6.6$\mu$m 
features in the 7 stars that have this feature are consistent with the temperature 
required to fit the $>$ 10--34 $\mu$m spectrum of AA Tau, 575\,K, shown by 
\citet{cana08}.  The water vapor in the circumstellar disk of AA Tau was 
inferred by \citet{cana08} to arise from the inner regions of its disk ($<$ 3 AU). 
We similarly suggest the water vapor to be located in the inner regions of the 
protoplanetary disks whose spectra we show.

The H$_{2}$CO absorption exemplars, DL Tau, HN Tau, and IQ Tau, all seem 
to require 2 temperatures of H$_{2}$CO to fit their 5.7 $\mu$m absorption bands.  
As noted previously, the cooler 50\,K component is added to ``fill in'' absorption 
at the center of the band that the warmer 500\,K component could not do by itself.  
In reality, the H$_{2}$CO probably spans a range of temperatures in these disks, 
but we opt for simplicity in our modeling.  The same is likely true for HCOOH 
and H$_{2}$O in our wider sample.

One point of comparison for our work is to the model of molecular emission by 
\citet{cana11} from RW Aur A as seen in its high resolution {\it Spitzer}-IRS spectrum.  
\citet{cana11} obtain a water temperature of 600\,K, with a column density of water 
of 1.55$\times$10$^{18}$cm$^{-2}$, and an inferred radius of emitting region of 1.49 AU.  In 
our modeling, we determine a water temperature of 1100\,K, a column density of 
1.7$\times$10$^{15}$cm$^{-2}$, and an inferred radius of emitting region of $\sim$ 4.3 AU.  
We model water lines over 5--7.5 $\mu$m wavelength, which arise mostly from $\nu_{2}$ = 1--0 
transitions, while water lines at $\lambda >$ 10 $\mu$m modeled by \citet{cana11} arise 
more from rotational transitions within the ground state of water.  Thus, we likely probe 
different disk conditions.  The water vapor in our model of RW Aur A has a lower column 
density spread out over a larger range of radii, so perhaps the hotter water vapor is located 
higher up in the atmosphere.  Thermo-chemical modeling shows hot water vapor relatively 
high in the disk atmosphere \citep[$>$1000\,K gas located 1 AU above midplane at R = 
3-4 AU from the star; see Figure 7 of][]{woit09}.  However, it is 
more likely that the higher resolution spectra modeled by \citet{cana11} simply provide 
better constraints on the water vapor than the low resolution Short-Low 5--7.5 $\mu$m 
spectra we model.  The higher resolution {\it Spitzer}-IRS spectra begin to 
separate emission lines (though such spectra do not necessarily outright resolve 
them), while Short-Low over 5--7.5 $\mu$m does not begin to resolve water emission 
lines.

\citet{meij09} suggest that protoplanetary disks that have experienced greater 
settling of dust to disk midplane, and therefore have higher gas-to-dust ratios 
in the upper disk layers, will have increased line-to-continuum ratios of emission 
from water vapor at infrared wavelengths.  \citet{fur06b} compared spectral indices 
from radiative transfer models that include the effects of dust settling \citep[][]{dalessio06} 
to observed spectral indices for TTSs in Taurus/Auriga 
and find that the mid-infrared continuum colors 
in general become bluer when there is greater dust settling and the disks become 
flatter.  \citet{fur11} measured the colors between 6--13 and 13--31 $\mu$m using 
the indices n$_{6-13}$ and n$_{13-31}$, which we cite for our sample in Table 1.  
Our sample's n$_{6-13}$ indices range from -1.39 to -0.62, while our samples 
n$_{13-31}$ indices range from -1.09 to 0.03.  As seen in Figure 7 of 
\citet{fur11}, such indices are consistent with a large fraction of the Taurus/Auriga 
TTS population, and as seen in Figure 22 of \citet{fur11}, this range of indices indicates 
significant settling, consistent with the suggestion by \citet{meij09}.  However, as 
we discussed in the previous subsection, we do not see trends between 6.6 $\mu$m 
equivalent width and either of the continuum color indices, n$_{6-13}$ and n$_{13-31}$.  
Perhaps any such trend is weak and might be found if our sample included more 
flared disks (i.e., with more positive indices).

We investigated the critical densities of the water vapor lines in the 5--7.5 $\mu$m 
region to determine how sensitive they might be to the degree of disk flaring (i.e., 
from dust settling) in protoplanetary disks.  Using a list of energy levels for the ground 
(rotation-only) and first few vibrational states for both ortho- and para-water from 
\citet{fajo08}, we followed the methodology of \citet{wat07} and computed the sum 
of the Einstein A-coefficients of permitted downward radiative transitions and the sum 
of the collisional rate coefficients for H$_{2}$-H$_{2}$O \citep[also from][]{fajo08} for 
each possible level, and divided the former by the latter for each level.  We determined 
the allowed transitions for the first few vibrational levels of water using the selection 
rules described by \citet{thibik05}.  From the \citet{fajo08} collisional rate coefficients, we 
obtain critical densities of about 10$^{7}$ to 10$^{11}$ cm$^{-3}$ for the first 45 levels (i.e., 
the ground state) of both ortho- and para-water.  These values are somewhat lower 
than the 10$^{10}$ to 10$^{12}$ cm$^{-3}$ determined by \citet{wat07} for water using 
levels and collisional rate coefficients from \citet{green93} and \citet{phil96}.  \citet{fajo08} 
note that their lower rates may be in error by up to 1 or 2 orders of magnitude, which 
perhaps helps explain this discrepancy.  However, the \citet{fajo08} collisional rate 
coefficients have the advantage of covering a wide range of transitions, and, thus, 
self-consistency.  The lines in the 5--7.5 $\mu$m region giving rise, in large part, to the 
emission seen in our models that match the observed spectra are mostly $\nu_{2}$ = 
1--0 transitions, though a few $\nu_{2}$ = 2--1 transitions also contribute.  The critical 
densities of these lines assuming the \citet{fajo08} collisional rate coefficients are in 
the range of about 10$^{10}$ to 6$\times$10$^{12}$ cm$^{-3}$.  Thus, the critical densities 
giving rise to the 5--7.5 $\mu$m water emission are, on average, an order of magnitude 
or two higher than those giving rise to the pure rotational transitions of water seen 
longward of 10 $\mu$m wavelength, suggesting the 5--7.5 $\mu$m water lines arise 
from denser disk regions.  Thus, one might expect that the disks with least dust settling may hide 
their water vapor emission, assuming the scale height of the dust in the inner few 
AU of the disks increases sufficiently as a result of the increased flaring.

Our model of the FU Orionis star V1057 Cyg is interesting in that it lacks absorption from 
HCOOH or H$_{2}$CO, unlike the TTSs in our sample.  Its spectrum is fit well with 
absorption {\it only} from H$_{2}$O at 800\,K in front of a 950\,K blackbody of radius 1.1 AU, 
with additional continuum emission from a 200\,K blackbody of radius 8.4 AU.  
A glance at the spectra of other FU Orionis stars 
shown by \citet{green06} suggests they have absorption features in the 5--7.5 $\mu$m 
range largely or totally due to water vapor.  Both having water vapor absorption instead 
of emission and lacking HCOOH and H$_{2}$CO absorption seems to differentiate this 
small sample of FU Orionis stars and our TTS sample.  The water vapor absorption profiles of 
FU Orionis stars is likely due to internal heating from viscous accretion, while the water vapor 
emission profiles of TTS's is likely due to external heating by stellar irradiation.  The 5 FU 
Orionis stars studied by \citet{green06} are among the least extinguished (flat spectrum) FU 
Orionis stars by several different indices \citet{green13}.  Those with more extinction show absorption from various 
ices over the 5--7.5 $\mu$m range typical of Class 0/I protostars \citep[][]{green06,quanz07}.

\citet{qi13} detected formaldehyde in the TTS disk of TW Hya and in the disk 
of the Herbig Ae star HD 163296 at millimeter wavelengths using the Submillimeter 
Array (SMA).  They propose that formaldehyde may form from hydrogenation of CO ice.  
\citet{aikawa12} constructed psuedo-time-dependent models of molecules in a young 
protoplanetary disk, though it is not clear how these models perform concerning 
formaldehyde.  Formic acid has been detected before from star-forming regions 
at radio wavelengths \citep[][]{ikeda01,remijan06}.  \citet{aikawa12} also expect 
formic acid to play a significant role in disk chemistry.  One major formation pathway 
for HCOOH gas comes from dissociative recombination of CH$_{3}$O$_{2}^{+}$, 
which itself comes from a HCO$^{+}$+H$_{2}$O reaction \citep[][]{aikawa12}.  It is 
also worth mentioning that formaldehyde (H$_{2}$CO) and formic acid (HCOOH) 
are chemically very similar, exchanging an H in formaldehyde for an OH in formic 
acid.  Thus, it may be no coincidence that we see formaldehyde present in some TTS disks 
and formic acid in other TTS disks.  We further note that both formaldehyde and formic 
acid ices are suggested by {\it Spitzer}-IRS spectra of Class I/II young stellar objects in 
the Taurus-Auriga star-forming region \citep[][]{zas09}.  Formaldehyde was also found 
in absorption in 3.6 $\mu$m spectra of the protostar W33A \citep[][]{rou06}.  If these molecules are present 
in ices falling onto newly-formed protoplanetary disks, then they may also remain in the 
later phase when the envelope has finished falling onto the disk, for us to see in TTS spectra.

One possible explanation for the presence of formaldehyde gas in disks providing strong 
absorption is that the formaldehyde gas there is short-lived.  Formaldehyde is known to 
undergo unimolecular dissociation \citep[][]{troe07}.  The rate coefficient, k$_{\rm Mol,0}$, 
determined by \citet{troe07} for formaldehyde dissociating to molecular hydrogen and CO 
is only valid between the temperatures of 1200--3500\,K.  We estimate the lifetime of 
formaldehyde in the disk of FZ Tau.  Replacing the argon concentration in the 
\citet{troe07} formula with an estimate of the disk gas density in its upper disk layers from 
the critical density of 10$^{11}$ cm$^{-3}$ determined by \citet{wat07} for the disk around 
the protostar NGC 1333 IRAS 4B \citep[this value is in between the estimates of minimum 
and maximum gas density, n$_{\rm min}$ and n$_{0}$, respectively, at a radius of $\sim$ 
3 AU in the disk model of][]{cg97}, and assuming the disk gas is at the temperature of the 
H$_{2}$O (1200\,K; see Table 2), we compute a half-life (the time it takes half of the population of this 
molecule to dissociate into molecular hydrogen and CO) of formaldehyde in the FZ Tau 
disk of $\sim$ 240 years.  Such a short lifetime may provide a natural explanation for the 
general lower temperatures of formaldehyde in our models (Table 2) in that hotter 
formaldehyde molecules tend to dissociate and cease to exist, but we would advocate for 
more detailed follow-up calculations in the future.

\section{Conclusions}

We modeled the emission and absorption bands in the 5--7.5 $\mu$m region of 13 
TTS in the Taurus-Auriga star-forming region.  We show strong evidence for water 
vapor emission in the seven sources showing the 6.6$\mu$m emission feature.  Such 
water emission contrasts with the absorption from water vapor seen in the spectrum of 
the FU Orionis star V1057 Cyg.  It is remarkable how much of the spectral detail of the 
V1057 Cyg spectrum over 5--7.5 $\mu$m is matched by our model that includes 
absorption from water vapor at a single temperature of 800\,K (with blackbodies providing continuum).  
For the entire sample, there is also absorption centered near 5.6-5.7$\mu$m that is 
consistent for some of the stars with formaldehyde and for others with formic acid.  
However, these are not the only stars in the 
Taurus/Auriga TTS population that show evidence of water vapor emission 
at 6.6$\mu$m or the absorption around 5.6--5.7$\mu$m from formaldehyde and/or 
some other gas species.  Numerous {\it Spitzer}-IRS spectra of TTSs outside our 
present sample of 13 show a 6.6 $\mu$m emission feature and 5.7 $\mu$m 
absorption band \citep[e.g.,][]{mcc10,fur11,manoj11}.

For the TTS spectra requiring water vapor emission, the water vapor temperatures 
range between 600--1200\,K, the column densities range from 1.7$\times$10$^{15}$ cm$^{-2}$ 
to 3.6$\times$10$^{17}$ cm$^{-2}$, and the inferred radii of the emitting region range from 
200--1000 $\rsun$ (Table 2).  This results mostly in maximum optical depths for water 
lines between 5--7.5 $\mu$m mostly less than 1, except for DR Tau ($\tau_{max,1} \sim$ 
5; Table 2).  The cool blackbodies included to add longer wavelength continuum over 
this wavelength range go from 160--300\,K.

The TTS spectra requiring formaldehyde absorption have formaldehyde temperatures 
ranging from 50--1000\,K (sometimes they have formaldehyde at 2 temperatures), 
column densities ranging from 2$\times$10$^{17}$ cm$^{-2}$ to 1.3$\times$10$^{18}$ cm$^{-2}$, 
and inferred radii of the absorbing region ranging from 10--22 $\rsun$.  This results in 
formaldehyde 5--7.5 $\mu$m region maximum optical depths for any one component 
ranging from 2--11.  The temperatures for the blackbody underlying these formaldehyde 
slabs range from 1000--1400\,K; i.e., the hot inner regions of the disks (Table 2).

The TTS spectra requiring formic acid absorption have formic acid temperatures at 
either 500 or 1000\,K, column densities ranging from 4$\times$10$^{16}$ cm$^{-2}$ to 
3.5$\times$10$^{18}$ cm$^{-2}$, and inferred radii of the absorbing region ranging from 
7 to 19 $\rsun$.  Maximum optical depths over 5--7.5 $\mu$m for any formic acid 
line range from 0.2--2.5.  The blackbody underlying each formic acid slab ranges 
in temperature from 1300--1600\,K, which (as with the blackbodies underlying the 
formaldehyde slabs) suggests an origin in the hot inner disk regions.

To confirm the identifications of formaldehyde and formic acid from the observed 5.6--5.7 
$\mu$m absorption bands, we would suggest observations of nearby infrared bands of 
these molecules.  Formic acid has the $\nu_{6}$ band centered near 9.05 $\mu$m 
\citep[][]{bas06} of approximately 
the same strength as the 5.65 $\mu$m band, as is revealed by the HITRAN2008 line list 
for formic acid \citep[][]{roth09}.  There is a hint of shallow absorption near this wavelength 
as seen in the {\it Spitzer}-IRS spectrum of DF Tau \citep[][]{sarg09b}; however, this 
wavelength region is confused with emission from silicate dust.  Therefore, it would be 
desirable to obtain high resolution N-band spectroscopy within the 8.75--9.25 $\mu$m 
range (approximately the span of the HCOOH 9.05 $\mu$m band) of stars whose 
{\it Spitzer}-IRS 5--7.5 $\mu$m spectra show absorption bands centered at 5.65 $\mu$m 
that are well-modeled by formic acid.  Formic acid also possesses an additional band 
of comparable but slightly lesser strength centered near 15.6 $\mu$m \citep[a blend of 
the $\nu_{7}$ and $\nu_{9}$ bands; see][]{bas06}.  Lines have been computed for 
formic acid, however, at present, it appears that it is possible only to compute relative 
line intensities for this band \citep[][]{perrin02}.  There are also weaker bands of formic 
acid centered near 2.8 and 3.4 $\mu$m \citep[$\nu_{1}$ and $\nu_{2}$, 
respectively][]{bas06}, but we could not find any line data for them.  For formaldehyde, 
there is a band of comparable strength to the 5.7 $\mu$m band centered near 3.55 
$\mu$m.  This was the band whose lines were observed by \citet{rou06} in the infrared 
spectrum of the protostar W33A.  There are no other infrared bands of comparable strength 
to these two bands for formaldehyde, so we would advocate high resolution L-band 
($\sim$ 3.55 $\mu$m) observations of stars whose {\it Spitzer}-IRS spectra show a 5.7 
$\mu$m absorption band that is best modeled by formaldehyde, to confirm the presence 
of formaldehyde based on the 5.7 $\mu$m band.

Spectral data at 5--7.5$\mu$m 
will be quite powerful when combined with data on water vapor and other gases 
longward of 10$\mu$m, like what \citet{cana08} showed and modeled for AA 
Tau.  Additionally, data on water vapor, when combined with data on CO 
\citep[see][]{naj96,naj03}, OH \citep{mand08}, etc., promises to reveal much 
about the physics and chemistry of gases in protoplanetary disks.  
Higher-resolution spectroscopic follow-up by current missions, such as SOFIA, and 
future missions such as {\it James Webb Space Telescope} (JWST) should begin to 
separate the lines of water vapor, formaldehyde, formic acid, and other gas 
species in the 5--7.5$\mu$m region of these stars' spectra.  
Using these observations, detailed modeling should 
yield valuable clues regarding the origin of water in the inner regions of 
protoplanetary disks, which is relevant to studies of the origin of water on planets 
in the habitable zones of stars.

\acknowledgements  This manuscript was supported by and is dedicated 
to the efforts of Paola D'Alessio, who will be sorely missed.  This work is 
based on observations made with 
the {\it Spitzer Space Telescope}, which is operated by the Jet Propulsion 
Laboratory, California Institute of Technology under NASA contract 1407.  
The authors wish to thank the referee, Geoffrey Blake, for comments that 
greatly improved this manuscript.  
We would like to thank Michael Mumma, Casey Lisse, Yuri Aikawa, and 
Karin {\"O}berg for helpful discussions.  
This research has made use of the SIMBAD database,
operated at CDS, Strasbourg, France.

\clearpage

\begin{figure}[t] 
  \epsscale{0.8}
  \plotone{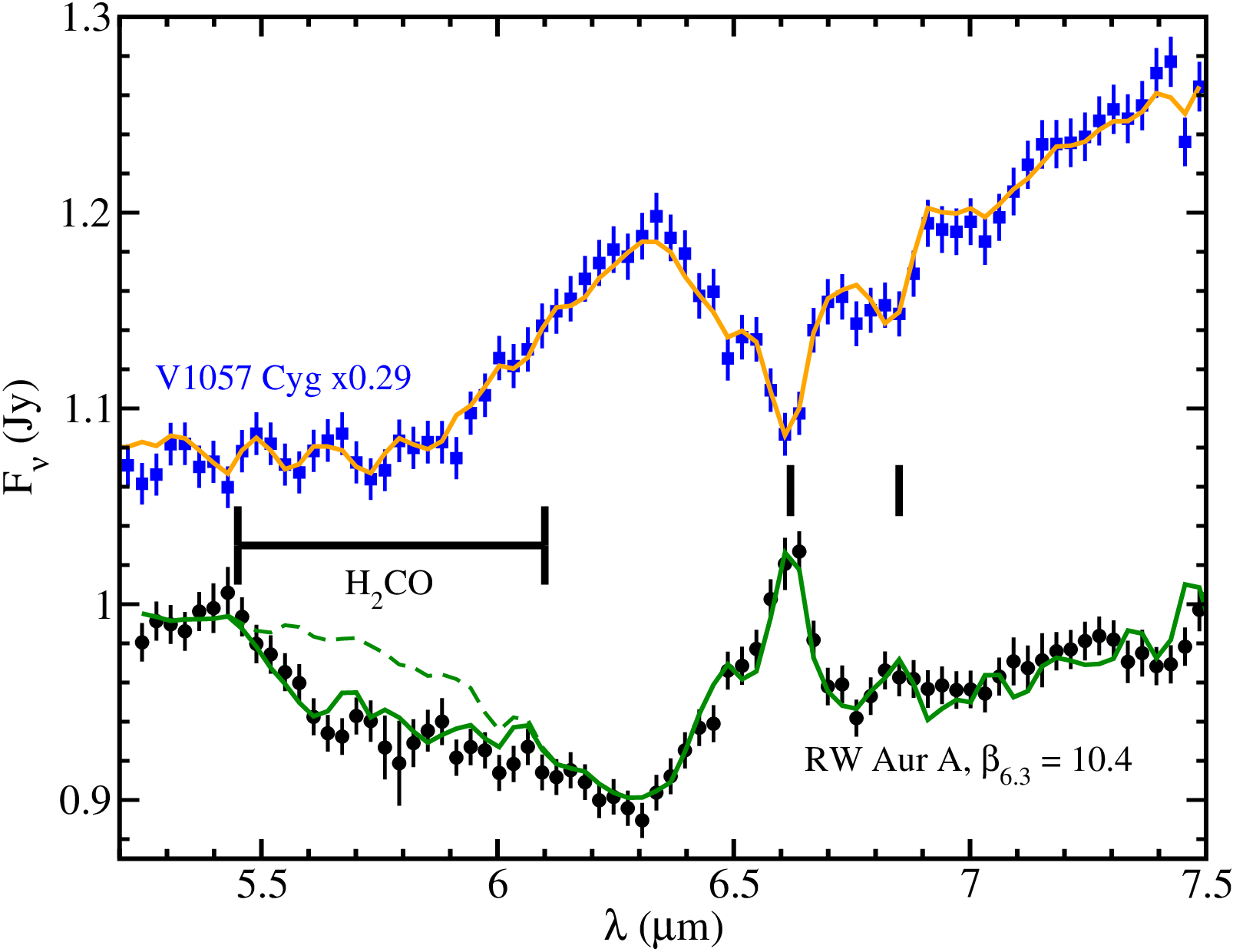}
  \caption[rwaura]{Detailed model of the spectra of RW Aur A (a TTS) and V1057 Cyg (an FU Ori star).  
  The observed spectrum of RW Aur A is the black filled circles with one sigma error bars.  The green line is the 
  model for RW Aur A.  The blue filled squares and associated one sigma error bars are the spectrum of V1057 
  Cyg, and the orange solid line is the model of this spectrum.  Both model and spectrum for V1057 Cyg are 
  scaled by 0.29.  For details, see Section 3.1 of the text.  The two short vertical black lines indicate the central 
  wavelengths of the 6.6 and 6.85 $\mu$m emission features discussed in the text.  The dashed line is a model 
  of RW Aur A that is the same as the solid line model, except that there is no formaldehyde absorption.  The 
  horizontal bar with vertical markers labeled H$_{2}$CO indicates the wavelength span of the formaldehyde 
  band.}
\end{figure}

\clearpage

\begin{figure}[t] 
  \epsscale{0.8}
  \plotone{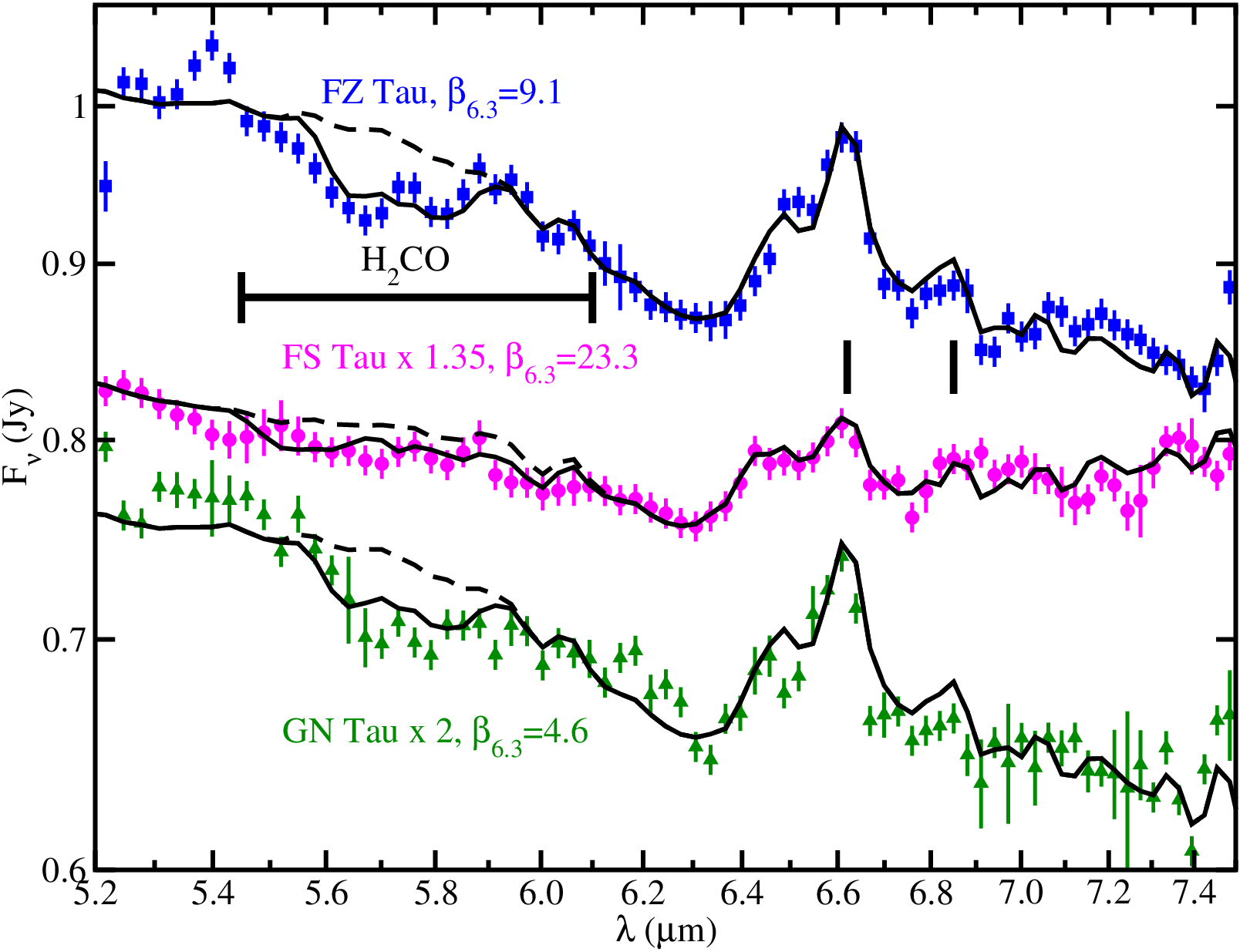}
  \caption[emh2co]{Spectra like RW Aur A, which show the 6.6$\mu$m water vapor emission feature but 
  require H$_{2}$CO in the model.  The data are solid points with one sigma error bars.  
  FZ Tau is blue filled squares, FS Tau is magenta filled circles, and GN Tau is green 
  upward-pointing triangles.  The name of the star is indicated by the label of the 
  same color as its spectrum.  The black line going through each spectrum is the model 
  of that spectrum.  The amount by which the observed spectrum is scaled 
  is indicated in the label for each source.  The label also gives the ratio of the flux at 
  6.3$\mu$m in the observed spectrum to that inferred from the photosphere plotted 
  for the source by \citet{fur06b}, $\beta_{6.3}$.  The two short vertical black lines have the same 
  meaning as in Figure 1, as do the dashed lines and horizontal bar with vertical markers.}
\end{figure}

\clearpage

\begin{figure}[t] 
  \epsscale{0.8}
  \plotone{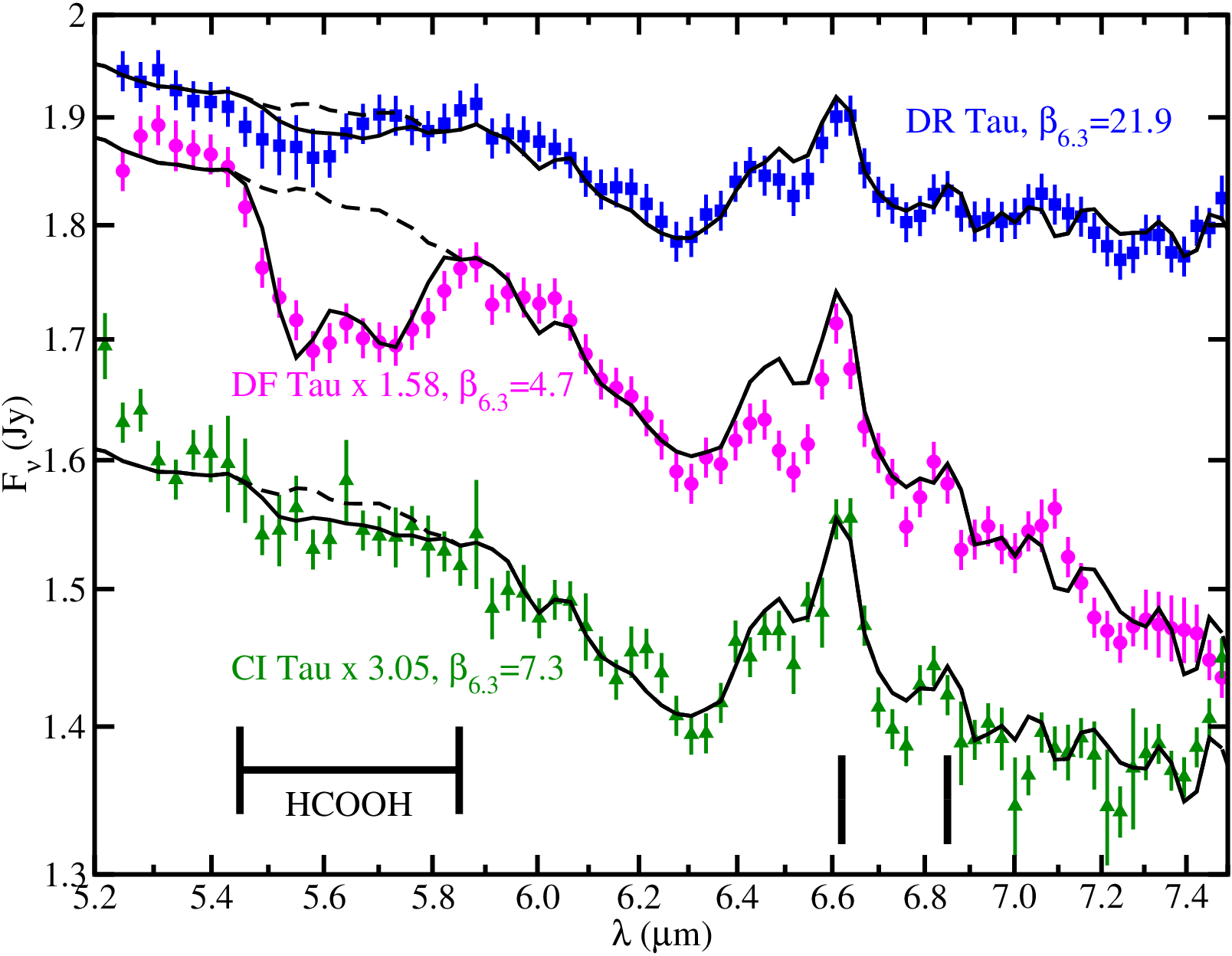}
  \caption[emhcooh]{Spectra showing the 6.6 $\mu$m water vapor emission feature but 
  requiring HCOOH in the model.  Same plot convention as in Figure 2, except that DR Tau 
  is the blue filled squares, DF Tau is the magenta filled circles, CI Tau is the green filled 
  upward-pointing triangles, and the horizontal bar with vertical markers now indicates the 
  wavelength span of the formic acid band.}
\end{figure}

\clearpage

\begin{figure}[t] 
  \epsscale{0.8}
  \plotone{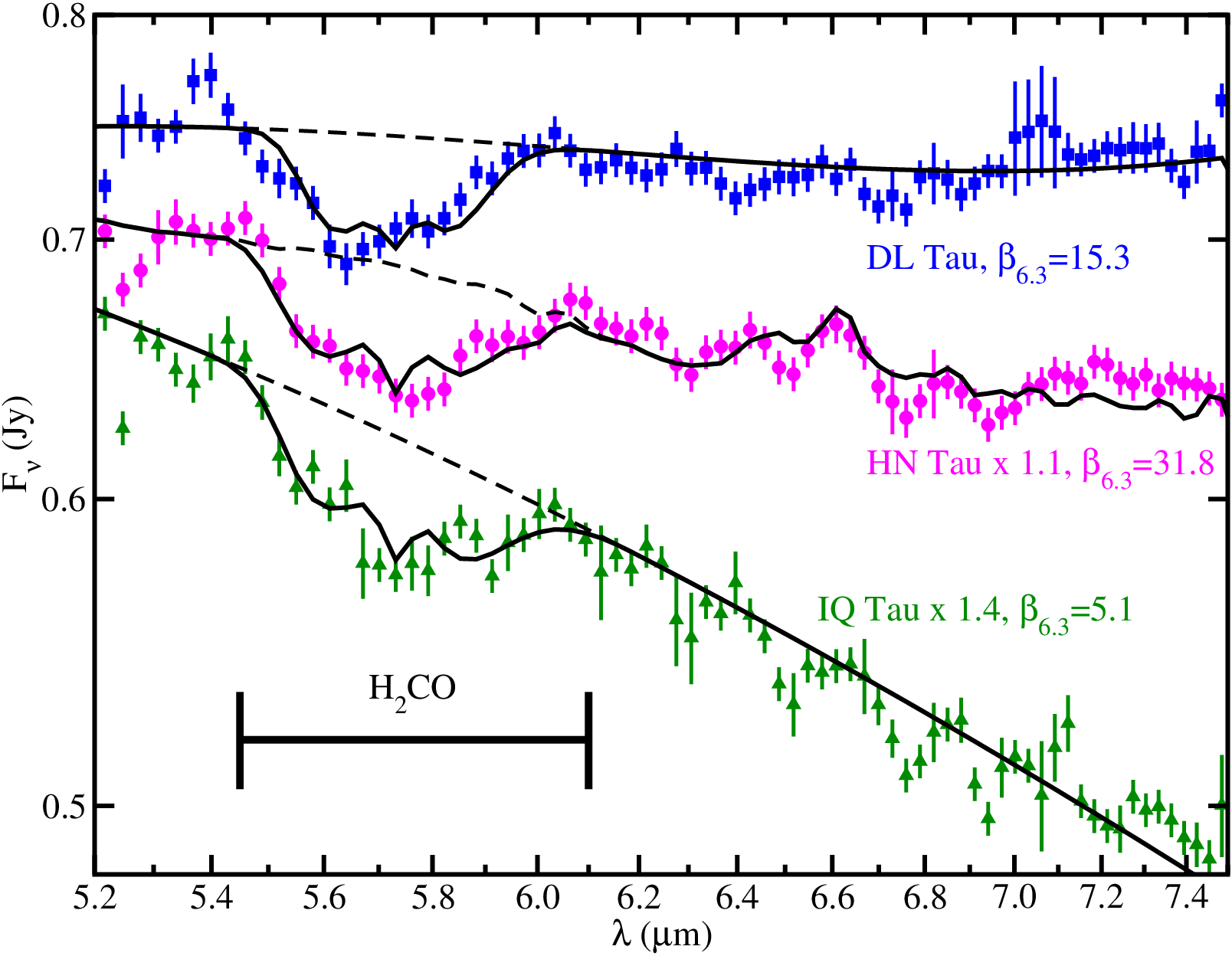}
  \caption[absh2co]{Spectra showing the 5.7 $\mu$m absorption band, modeled using H$_{2}$CO.  
  Same plot convention as in Figure 2, except that DL Tau is the blue filled squares, 
  HN Tau is the magenta filled circles, IQ Tau is the green filled upward-pointing triangles, and 
  the horizontal bar with vertical markers now indicates the wavelength span of the formaldehyde band.}
\end{figure}

\clearpage

\begin{figure}[t] 
  \epsscale{0.8}
  \plotone{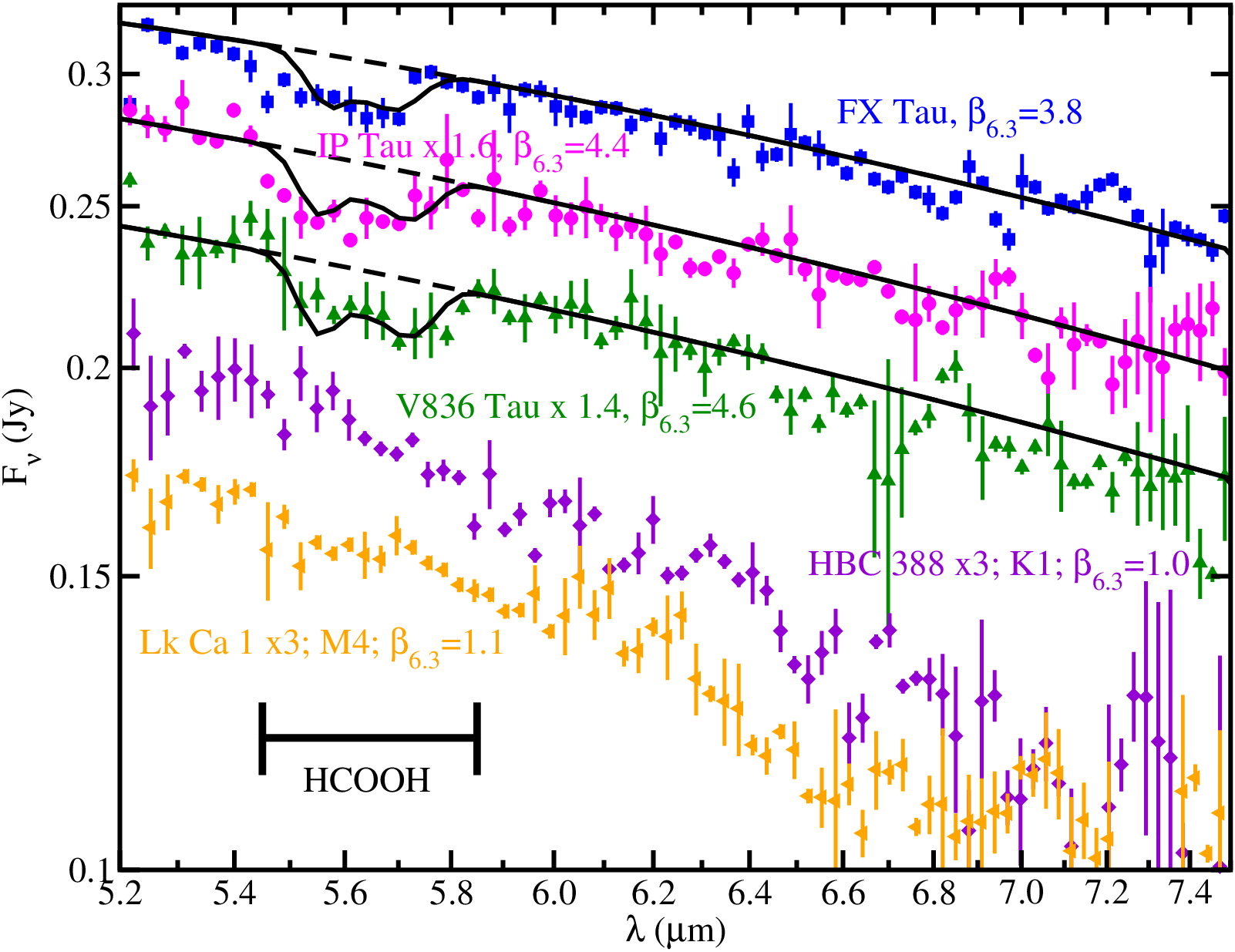}
  \caption[abshcooh]{Spectra showing the 5.7 $\mu$m absorption band, modeled using HCOOH.  
  Same plot convention as in Figure 2, except that FX Tau is the blue filled squares, IP Tau 
  is the magenta filled circles, V836 Tau is the green upward-pointing filled triangles, and 
  the horizontal bar with vertical markers now indicates the wavelength span of the formic acid 
  band.  Also shown are 
  the Class III YSO sources HBC 388 as violet filled diamonds and Lk Ca 1 as orange filled 
  leftward-pointing triangles.}
\end{figure}

\clearpage

\begin{figure}[t] 
  \epsscale{0.8}
  \plotone{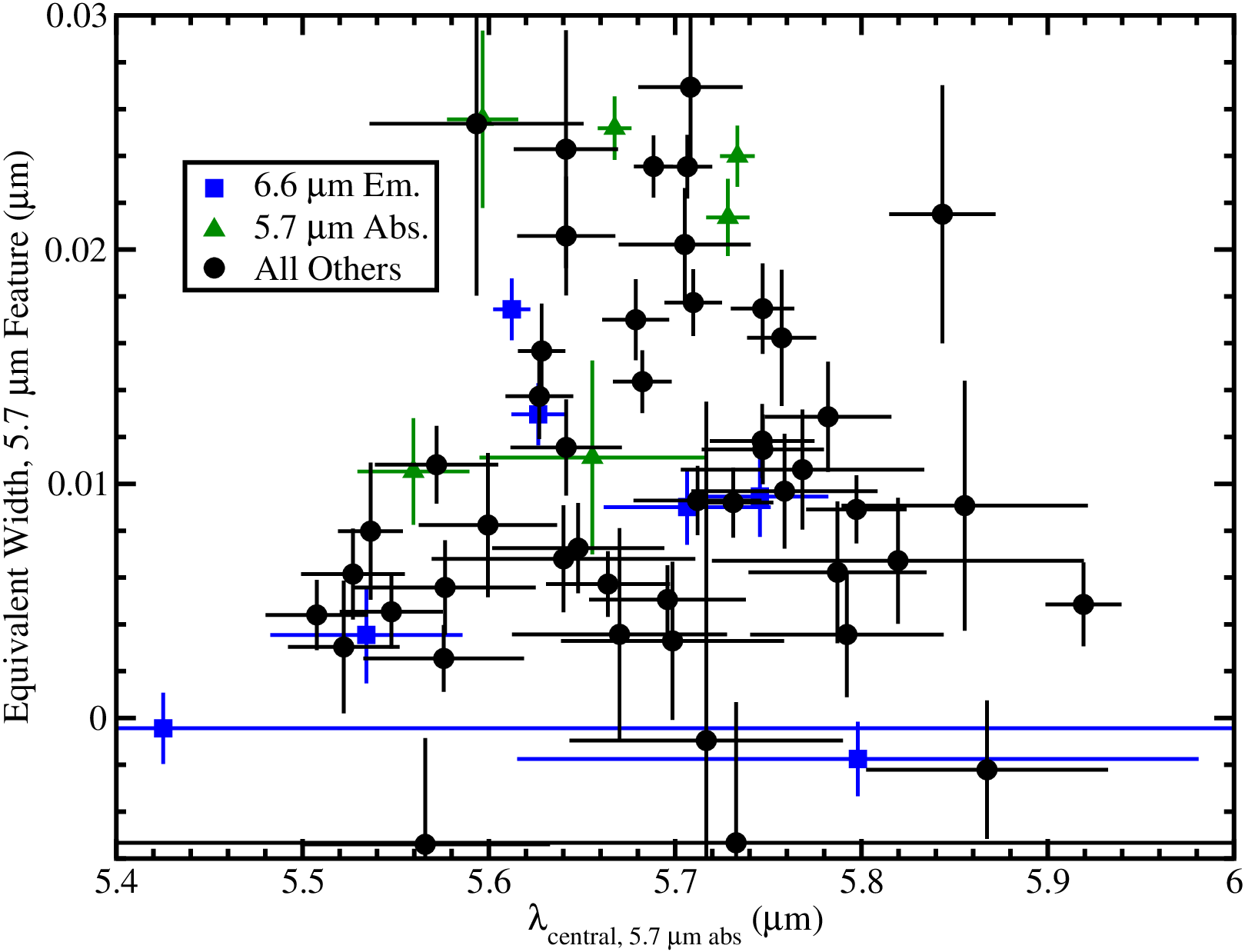}
  \caption[ew57vslamcen57]{Plot of the measured 5.7 $\mu$m absorption feature equivalent 
  width (in $\mu$m) versus the central wavelength of the 5.7 $\mu$m absorption feature.  The 
  blue filled squares are the 7 spectra in our sample showing strong 6.6 $\mu$m emission features, 
  the green filled triangles are the 6 spectra in our sample showing strong 5.7 $\mu$m absorption 
  features, and the black filled circles are the other 48 TTSs in the \citet{sarg09b} sample (not including 
  CoKu Tau/4, DM Tau, GM Aur, and Lk Ca 15 because they are transitional disks) from the 
  Taurus/Auriga star-forming region.}
\end{figure}

\clearpage

\begin{figure}[t] 
  \epsscale{0.8}
  \plotone{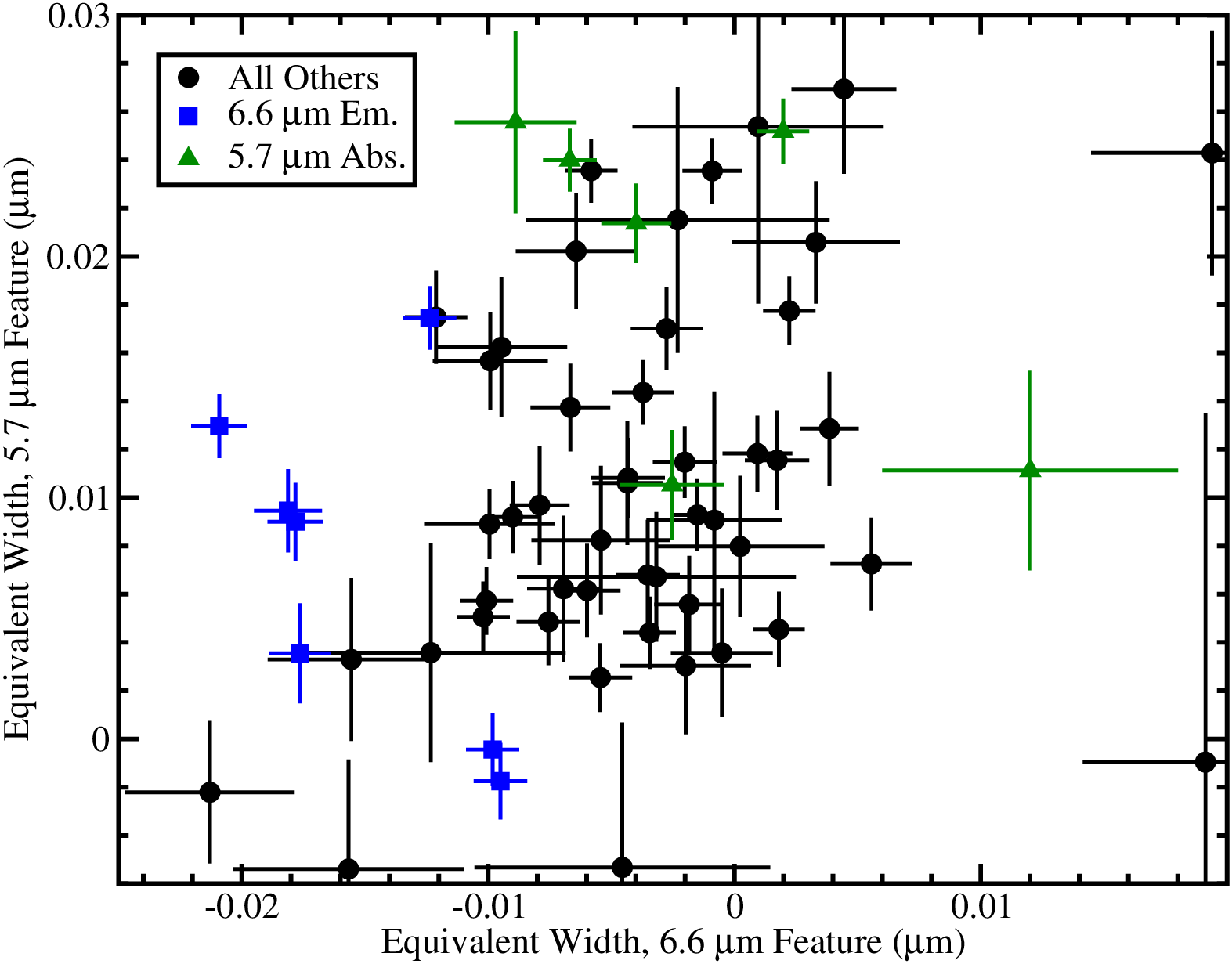}
  \caption[ew57vsew66]{Plot of the measured 5.7 $\mu$m absorption feature equivalent width (in 
  $\mu$m) versus the measured 6.6 $\mu$m emission feature equivalent width (in $\mu$m).  Same 
  plot style as in Figure 6.}
\end{figure}

\clearpage

\begin{figure}[t] 
  \epsscale{0.8}
  \plotone{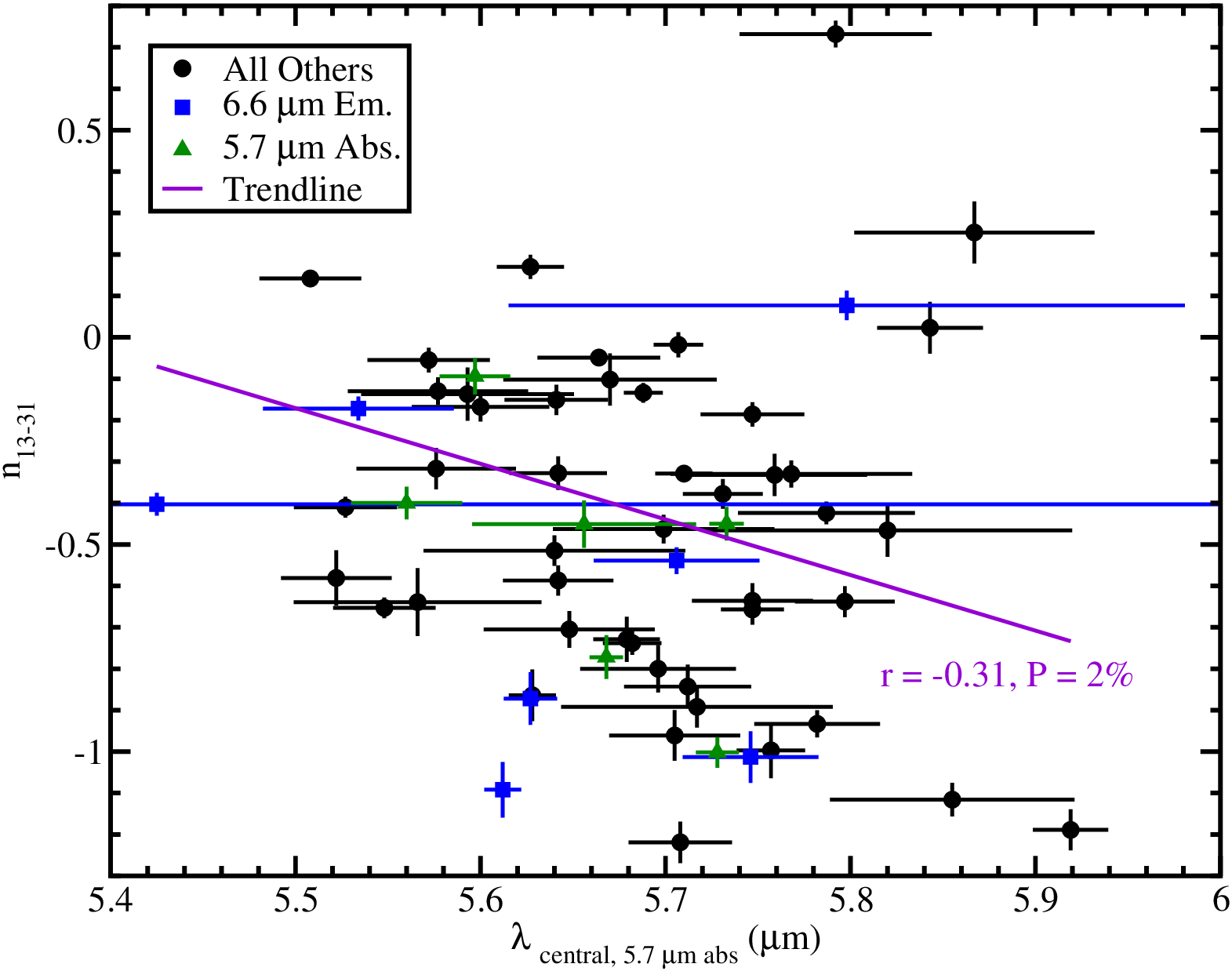}
  \caption[n1331vslamcen57]{Plot of the 13-to-31 $\mu$m index versus the 5.7 $\mu$m absorption 
  feature central wavelength.  The trend line determined from least-squares fitting of all 59 plotted 
  points is the violet line, and the correlation coefficient and probability of a correlation coefficient 
  of equal or greater magnitude being found for a non-correlated data set \citep[see][]{sarg09b} are 
  indicated beside the line.  Otherwise, same plot style as in Figure 6.}
\end{figure}

\clearpage

\begin{figure}[t] 
  \epsscale{0.8}
  \plotone{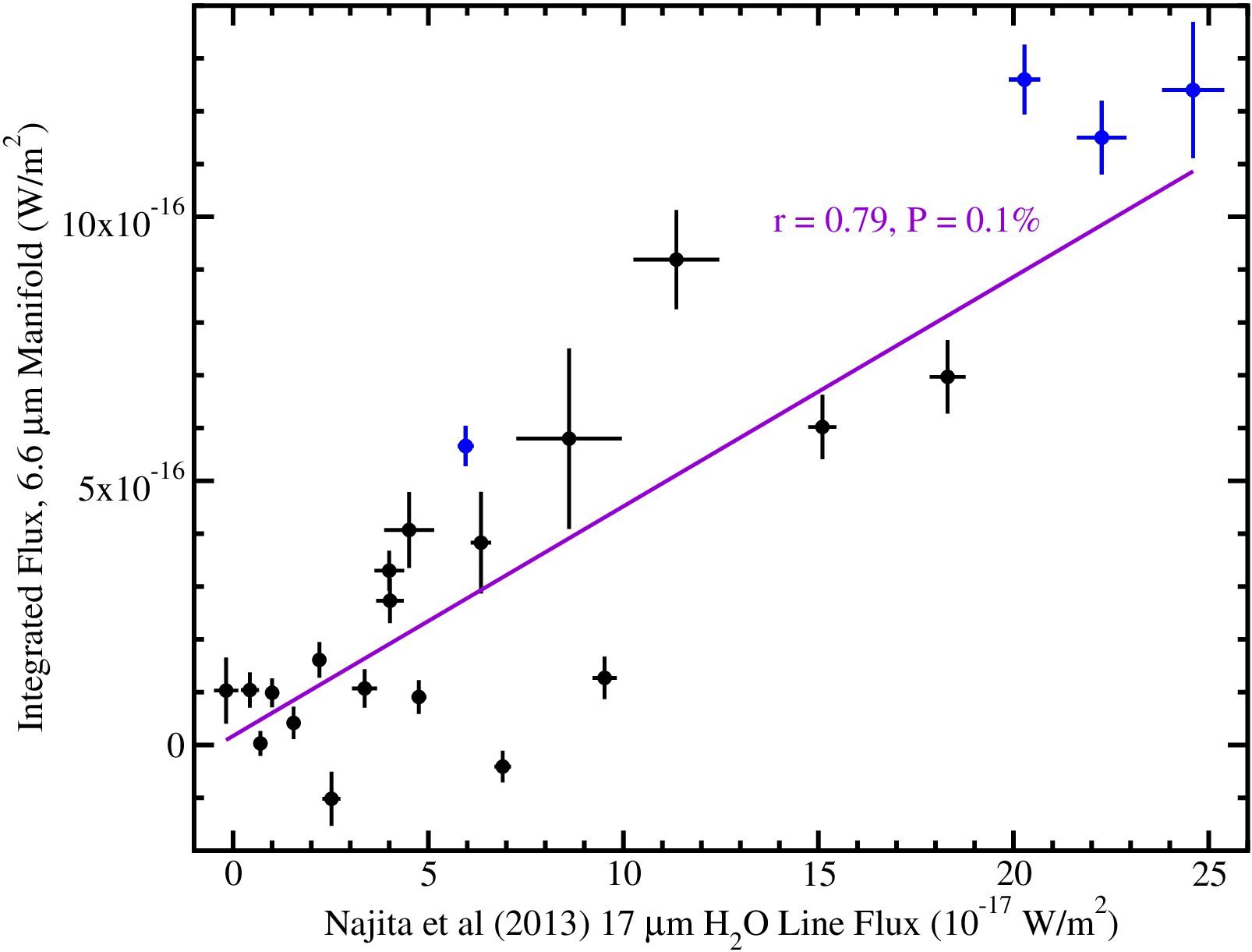}
  \caption[f66vsf17]{Plot of the 17 $\mu$m H$_{2}$O line flux (in 10$^{-17}$ W m$^{-2}$) from 
  \citet{naj13} versus 6.6 $\mu$m 
  emission manifold integrated flux (in W m$^{-2}$).  Blue points are CI Tau, DR Tau, FZ Tau, and RW 
  Aur A, as discussed in the text, while the black points are all the other stars in common between the 
  \citet{naj13} sample and the sample for which we measured 6.6 and 5.7 $\mu$m equivalent widths 
  and integrated fluxes.  As with Figure 8, the violet line is the trend line, and the correlation coefficient 
  and P are indicated beside the line.}
\end{figure}

\clearpage

\begin{landscape}
\begin{table}[h,t]
{
\caption[Analysis]{Sample\label{table1}}
\begin{tabular}{lccccccccccc}
\hline \hline
          
          & Sp.
          & 
          & 
          & Misptg.
          & Misptg.
          & 5.7$\mu$m
          & 5.7$\mu$m
          & 5.7$\mu$m
          & 6.6$\mu$m
          & 6.6$\mu$m
          & 6.6$\mu$m\\
          Star
          & Ty.$^{\rm a}$
          & n$_{6-13}^{\rm a}$
          & n$_{13-31}^{\rm a}$
          & SL2n1$^{b}$
          & SL2n2$^{b}$
          & $\lambda_{central}^{c}$
          & EW$\times$1000$^{d}$
          & $\left|EW\right|$/$\sigma_{EW}^{e}$
          & $\lambda_{central}^{c}$
          & EW$\times$1000$^{d}$
          & $\left|EW\right|$/$\sigma_{EW}^{e}$\\
\hline
\multicolumn{12}{c}{6.6 $\mu$m Emission$^{f}$} \\
CI Tau & K7 & -0.97 & -0.17 & 0.062 & -0.096 & 5.534$\pm$0.052 & 3.55$\pm$2.1 & 1.7 & 6.573$\pm$0.012 & -17.6$\pm$1.2 & 15\\
DF Tau & M2 & -1.39 & -1.09 & \nodata & \nodata & 5.612$\pm$0.010 & 17.5$\pm$1.3 & 13 & 6.597$\pm$0.007 & -12.4$\pm$1.1 & 11\\
DR Tau & K5 & -0.90 & -0.40 & \nodata & \nodata & 5.425$\pm$1.529 & -0.4$\pm$1.5 & 0.27 & 6.574$\pm$0.014 & -9.8$\pm$1.1 & 8.9\\
FS Tau & M0 & -0.62 & 0.03 & -0.075 & 0.096 & 5.798$\pm$0.183 & -1.7$\pm$1.6 & 1.1 & 6.529$\pm$0.023 & -9.5$\pm$1.1 & 8.6\\
FZ Tau & M0 & -1.18 & -0.89 & 0.363 & 0.768 & 5.627$\pm$0.014 & 13.0$\pm$1.3 & 10 & 6.582$\pm$0.006 & -20.9$\pm$1.1 & 19\\
GN Tau & M2.5 & -0.96 & -1.02 & \nodata & \nodata & 5.746$\pm$0.037 & 9.5$\pm$1.7 & 5.6 & 6.572$\pm$0.007 & -18.1$\pm$1.4 & 13\\
RW Aur A & K3 & -0.69 & -0.54 & \nodata & \nodata & 5.706$\pm$0.045 & 9.0$\pm$1.6 & 5.6 & 6.580$\pm$0.007 & -17.8$\pm$1.1 & 16\\
\hline
\multicolumn{12}{c}{5.7 $\mu$m Absorption$^{f}$} \\
DL Tau & K7 & -0.77 & -0.77 & -0.662 & -0.278 & 5.668$\pm$0.009 & 25.2$\pm$1.4 & 18 & 6.425$\pm$0.044 & 2.0$\pm$1.1 & 1.8\\
FX Tau & M1 & -1.07 & -0.40 & 0.095 & 0.315 & 5.560$\pm$0.030 & 10.5$\pm$2.3 & 4.6 & 6.551$\pm$0.063 & -2.5$\pm$2.1 & 1.2\\
HN Tau & K5 & -0.70 & -0.44 & \nodata & \nodata & 5.733$\pm$0.009 & 24.0$\pm$1.3 & 18 & 6.572$\pm$0.018 & -6.7$\pm$1.1 & 6.1\\
IP Tau & M0 & -0.93 & -0.11 & 0.335 & 0.132 & 5.597$\pm$0.019 & 25.6$\pm$3.8 & 6.7 & 6.481$\pm$0.021 & -8.9$\pm$2.5 & 3.6\\
IQ Tau & M0.5 & -1.30 & -1.00 & -0.248 & -0.61 & 5.728$\pm$0.012 & 21.4$\pm$1.6 & 13 & 6.528$\pm$0.047 & -4.0$\pm$1.4 & 2.9\\
V836 Tau & K7 & -1.38 & -0.45 & \nodata & \nodata & 5.656$\pm$0.061 & 11.1$\pm$4.1 & 2.7 & 6.662$\pm$0.031 & 12.0$\pm$6.0 & 2.0\\
\hline
\end{tabular}
\tablenotetext{a}{\footnotesize The spectral type (Sp. Ty.) and spectral 
continuum indices n$_{6-13}$ and n$_{13-31}$ for our sample come 
from \citet{fur11}.}
\tablenotetext{b}{\footnotesize Mispointing in the cross-dispersion 
direction (i.e., along the slit length), measured in pixels.  ``SL2nX'' means 
Short-Low order 2, nod X, where X is either 1 or 2.  For mapping-mode 
observations \citep[those stars observed in 2$\times$3 rasters; see][]{sarg06}, 
the (row 2, column 2) position is treated as nod 1, and the (row 1, column 2) 
position is treated as nod 2.}
\tablenotetext{c}{\footnotesize The central wavelength for the feature or band, 
as defined in the text, in microns.}
\tablenotetext{d}{\footnotesize The equivalent width (EW) of the feature or band, 
as defined in the text, in microns, multiplied by 1000.}
\tablenotetext{e}{\footnotesize Significance of the feature, defined as the ratio of the 
equivalent width to its uncertainty.}
\tablenotetext{f}{\footnotesize The labels ``6.6 $\mu$m Emission'' 
and ``5.7 $\mu$m Absorption'' indicate whether the star in the sample is a 6.6 $\mu$m 
emission feature exemplar or a 5.6--5.7 $\mu$m absorption band exemplar, respectively.}
}
\end{table}
\end{landscape}

\clearpage

\begin{landscape}
\begin{table}[h,t]
{
\caption[Analysis]{Models\label{table2}}
\begin{tabular}{lccccccccccc}
\hline \hline
\multicolumn{6}{c}{H$_{2}$O} & \multicolumn{6}{c}{H$_{2}$CO or HCOOH}\\
          
          & T$_{1}$
          & N$_{1}$
          & T$_{bb1}$
          & R$_{bb1}^{a}$
          & 
          & T$_{2}$
          & N$_{2}$
          & T$_{bb2}$
          & R$_{bb2}^{a}$
          & 
          & abund.\\
          Star
          & (K)
          & (10$^{18}$ cm$^{-2}$)
          & (K)
          & (AU)
          & $\tau_{max,1}$
          & (K)
          & (10$^{18}$ cm$^{-2}$)
          & (K)
          & ($\rsun$)
          & $\tau_{max,2}$
          & ratio$^{b}$\\
\hline
\multicolumn{12}{c}{H$_{2}$O and H$_{2}$CO} \\
DL Tau$^{c}$ & \nodata & \nodata & 200 & 2.0 & \nodata & 300,50 & 1.1,0.2 & 1000 & 22 & 10,7.0 & 0\\
FS Tau & 600 & 0.01 & 200 & 2.5 & 0.13 & 1000 & 1 & 1400 & 13 & 2.3 & 18\\
FZ Tau & 1200 & 0.005 & 190 & 2.5 & 0.026 & 150 & 0.8 & 1400 & 16 & 14 & 6.7\\
GN Tau & 1100 & 0.01 & 200 & 1.2 & 0.058 & 200 & 0.5 & 1400 & 10 & 6.7 & 13\\
HN Tau$^{c}$ & 900 & 0.0035 & 200 & 1.8 & 0.026 & 500,50 & 1.3,0.3 & 1200 & 16 & 8.2,11 & 1.4\\
IQ Tau$^{c}$ & \nodata & \nodata & \nodata & \nodata & \nodata & 500,50 & 1.2,0.2 & 1400 & 11 & 7.5,7.0 & 0\\
RW Aur A & 1100 & 0.0017 & 190 & 4.3 & 0.0099 & 500 & 0.7 & 1400 & 16 & 4.4 & 7.9\\
\multicolumn{12}{c}{H$_{2}$O and HCOOH} \\
CI Tau & 900 & 0.0085 & 200 & 1.6 & 0.062 & 1000 & 0.5 & 1400 & 12 & 0.35 & 15\\
DF Tau & 900 & 0.0022 & 160 & 4.7 & 0.016 & 1000 & 3 & 1400 & 18 & 2.1 & 2.3\\
DR Tau & 600 & 0.36 & 300 & 0.93 & 4.6 & 500 & 0.04 & 1600 & 19 & 0.22 & 980\\
FX Tau & \nodata & \nodata & \nodata & \nodata & \nodata & 500 & 0.2 & 1300 & 10 & 1.1 & 0\\
IP Tau & \nodata & \nodata & \nodata & \nodata & \nodata & 1000 & 2.8 & 1400 & 6.9 & 2.0 & 0\\
V836 Tau & \nodata & \nodata & \nodata & \nodata & \nodata & 1000 & 3.5 & 1400 & 6.8 & 2.5 & 0\\
\multicolumn{12}{c}{H$_{2}$O Only} \\
V1057 Cyg$^{d}$ & 800 & 60 & 200 & 8.4 & 74 & \nodata & \nodata & \nodata & \nodata & \nodata & $\infty$\\
\hline
\end{tabular}
\tablenotetext{a}{\footnotesize The distance, D, assumed for our 13 TTS from Taurus-Auriga is 139 
	pc \citep[][]{bert99}, so that the solid angle, $\Omega$, is $\pi$R$_{bb}^{2}$/D$^{2}$.}
\tablenotetext{b}{\footnotesize The ratio of the abundance of water in the model to the abundance of 
the ``other'' species (H$_{2}$CO or HCOOH) in the model, computed as a$_{H_{2}O}$/a$_{other}$, 
where a$_{H_{2}O}$ = N$_{1}\pi$R$_{bb1}^{2}$ and a$_{other}$ = N$_{2}\pi$R$_{bb2}^{2}$.}
\tablenotetext{c}{\footnotesize The model with the non-water gas species has two temperatures, 
the cooler gas slab on top of the warmer gas slab.}
\tablenotetext{d}{\footnotesize The distance assumed for the FU Orionis object V1057 Cyg is 600 pc 
(see text).  Also, as the model for this star is different from those of the TTSs in our sample, we present 
its parameters differently.  The blackbody behind the 800\,K water vapor is at 950\,K, is 1.1 AU in 
radius.  The 200\,K blackbody of 8.4 AU radius is a separate component in the V1057 Cyg model.}
}
\end{table}
\end{landscape}

\end{document}